\documentclass[10pt,nc,nature,notitlepage,superscriptaddress,twocolumn]{revtex4-2}
\usepackage{upgreek}
\usepackage{amsmath}
\usepackage{amssymb}
\usepackage{amsfonts}
\usepackage{bbm}
\usepackage{bm}
\usepackage{graphicx}
\usepackage{color}
\definecolor{dred}{rgb}{.8,0.2,.2}
\definecolor{ddred}{rgb}{.8,0.5,.5}
\definecolor{dblue}{rgb}{.2,0.2,.8}
\definecolor{dgreen}{rgb}{.2,0.5,.2}

\usepackage{dcolumn}
\usepackage{MnSymbol}
\usepackage{times}
\usepackage{verbatim}

\newcommand{\ket}[1]{|{#1}\rangle}

\newcommand{\bra}[1]{\langle{#1}|}

\usepackage[utf8]{inputenc}
\usepackage[T1]{fontenc}
\usepackage[english]{babel}  

\usepackage{tikz}

\usepackage{colortbl}
\usepackage{tabu,diagbox}
\usepackage{natbib}
\setcitestyle{super}

\usepackage{times}
\usepackage{dsfont}

\usepackage{amssymb,amsmath,amsfonts,amsthm}
\usepackage{mathtools}
\usepackage{verbatim}
\usepackage{enumerate}
\usepackage{adjustbox}
\usepackage{graphicx}
\graphicspath{figs/}
\usepackage[colorlinks,linkcolor=blue,anchorcolor=blue,urlcolor=blue,citecolor=blue]{hyperref}
\usepackage[capitalise]{cleveref}
\usepackage{bbm}
\usepackage{booktabs}

\usepackage{color}
\definecolor{dred}{rgb}{.8,0.2,.2}
\definecolor{ddred}{rgb}{.8,0.5,.5}
\definecolor{dblue}{rgb}{.2,0.2,.8}
\definecolor{dgreen}{rgb}{.2,0.5,.2}

\newcommand{\be}{\begin{equation}}
\newcommand{\ee}{\end{equation}}
\newcommand{\bea}{\begin{eqnarray}}
\newcommand{\eea}{\end{eqnarray}}

\makeatletter\renewcommand{\@biblabel}[1]{#1.}\makeatother

\usepackage[small,raggedright]{titlesec}
\titleformat*{\section} {\small\bf}
\titlespacing*{\section} {0pt}{10pt}{0pt}
\titlespacing*{\subsection} {0pt}{10pt}{0pt}
\begin{document}

\title{Experimental quantum simulation of non-Hermitian dynamical topological states using stochastic Schr\"{o}dinger equation}

\author{Zidong Lin}
\thanks{These authors contributed equally to this work.}
\affiliation{Shenzhen Institute for Quantum Science and Engineering and Department of Physics, Southern University of Science and Technology, Shenzhen 518055, China}

\author{Lin Zhang}
\thanks{These authors contributed equally to this work.}
\affiliation{International Center for Quantum Materials and School of Physics, Peking University, Beijing 100871, China}
\affiliation{Collaborative Innovation Center of Quantum Matter, Beijing 100871, China}
\affiliation{ICFO-Institut de de Ci\`encies Fot\`oniques, The Barcelona Institute of Science and Technology, Av. Carl Friedrich Gauss 3, 08860 Castelldefels (Barcelona), Spain}

\author{Xinyue Long}
\thanks{These authors contributed equally to this work.}
\affiliation{Shenzhen Institute for Quantum Science and Engineering and Department of Physics, Southern University of Science and Technology, Shenzhen 518055, China}

\author{Yu-ang Fan}
\affiliation{Shenzhen Institute for Quantum Science and Engineering and Department of Physics, Southern University of Science and Technology, Shenzhen 518055, China}

\author{Yishan Li}
\affiliation{Shenzhen Institute for Quantum Science and Engineering and Department of Physics, Southern University of Science and Technology, Shenzhen 518055, China}

\author{Kai Tang}
\affiliation{Shenzhen Institute for Quantum Science and Engineering and Department of Physics, Southern University of Science and Technology, Shenzhen 518055, China}

\author{Jun Li}
\affiliation{Shenzhen Institute for Quantum Science and Engineering and Department of Physics, Southern University of Science and Technology, Shenzhen 518055, China}
\affiliation{International Quantum Academy, Shenzhen, 518055, China}
\affiliation{Guangdong Provincial Key Laboratory of Quantum Science and Engineering, Southern University of Science and Technology, Shenzhen 518055, China}

\author{Xinfang Nie}
\affiliation{Shenzhen Institute for Quantum Science and Engineering and Department of Physics, Southern University of Science and Technology, Shenzhen 518055, China}
\affiliation{Guangdong Provincial Key Laboratory of Quantum Science and Engineering, Southern University of Science and Technology, Shenzhen 518055, China}

\author{Tao Xin}
\email{xint@sustech.edu.cn}
\affiliation{Shenzhen Institute for Quantum Science and Engineering and Department of Physics, Southern University of Science and Technology, Shenzhen 518055, China}
\affiliation{International Quantum Academy, Shenzhen, 518055, China}
\affiliation{Guangdong Provincial Key Laboratory of Quantum Science and Engineering, Southern University of Science and Technology, Shenzhen 518055, China}

\author{Xiong-Jun Liu}
\email{xiongjunliu@pku.edu.cn}
\affiliation{International Center for Quantum Materials and School of Physics, Peking University, Beijing 100871, China}
\affiliation{Collaborative Innovation Center of Quantum Matter, Beijing 100871, China}
\affiliation{International Quantum Academy, Shenzhen, 518055, China}

\author{Dawei Lu}
\email{ludw@sustech.edu.cn}
\affiliation{Shenzhen Institute for Quantum Science and Engineering and Department of Physics, Southern University of Science and Technology, Shenzhen 518055, China}
\affiliation{International Quantum Academy, Shenzhen, 518055, China}
\affiliation{Guangdong Provincial Key Laboratory of Quantum Science and Engineering, Southern University of Science and Technology, Shenzhen 518055, China}

\begin{abstract}
Noise is ubiquitous in real quantum systems, leading to non-Hermitian quantum dynamics, and may affect the fundamental states of matter. Here we report in experiment a quantum simulation of the two-dimensional non-Hermitian quantum anomalous Hall (QAH) model using the nuclear magnetic resonance processor. Unlike the usual experiments using auxiliary qubits, we develop a stochastic average approach based on the stochastic Schrödinger equation to realize the non-Hermitian dissipative quantum dynamics, which has advantages in saving the quantum simulation sources and simplifies implementation of quantum gates. We demonstrate the stability of dynamical topology against weak noise, and observe two types of dynamical topological transitions driven by strong noise. Moreover, a region that the emergent topology is always robust regardless of the noise strength is observed. Our work shows a feasible quantum simulation approach for dissipative quantum dynamics with stochastic Schrödinger equation and opens a route to investigate non-Hermitian dynamical topological physics.
\end{abstract}
\maketitle

\section*{INTRODUCTION}
As a fundamental notion beyond the celebrated Landau-Ginzburg-Wilson framework~\cite{landau1999statistical}, the topological quantum matter has stimulated extensive studies in recent years, with tremendous progress having been achieved in searching for various types of topological states~\cite{hasan2010colloquium,qi2011topological,chiu2016classification,ando2015topological,sato2017topological,xie2021higher}. A most important feature of topological matter is the bulk-surface correspondence~\cite{hasan2010colloquium,qi2011topological,chiu2016classification}, which relates the bulk topology to boundary states and provides the foundation of most experimental characterizations and observations of topological quantum phases, such as via transport measurements~\cite{Konig2007,Chang2013,He2017} and angle resolved photoemission spectroscopy~\cite{Hsieh2008,Chen2009,Xia2009}.

Despite the fact that topological phases are defined at the ground state at equilibrium, quantum quenches in recent studies provide nonequilibrium way to investigate topological physics~\cite{Caio2015,Vajna2015,Budich2016,Wilson2016,Gong2018,Flaschner2018,Song2018,Qiu2019,Huang2020,Unal2020,Hu2020,Hu2020Hopf,Lu2020,Wang2021,Mizoguchi2021}. Particularly, as a momentum-space counterpart of the bulk-boundary correspondence, the dynamical bulk-surface correspondence was proposed~\cite{Zhang2018,Zhang2019,Zhang2020Floquet,Yu2021,Li2021,Zhang2021AZ}, which relates the bulk topology of an equilibrium phase to nontrivial dynamical topological phase emerging on certain momentum subspaces called band-inversion surfaces (BISs) when quenching the system across topological transitions. This dynamical topology enables a broadly applicable way to characterize and detect topological phases by quantum dynamics, and has triggered many experimental studies in quantum simulations, such as in ultracold atoms~\cite{Sun2018,Yi2019}, nitrogen-vacancy defects in diamond~\cite{Wang2019,Ji2020,Chen2021}, nuclear magnetic resonance (NMR)~\cite{Xin2020}, and superconducting circuits~\cite{Niu2021}.

The quench induced dynamical topological phase has been mainly studied in Hermitian systems, while the system is generally non-Hermitian when coupled to environment~\cite{Breuer2007}. Recently, the interplay between non-Hermiticity and topology has attracted considerable attention~\cite{Ashida2020,Bergholtz2021}, with rich phenomena being uncovered, such as the exotic topological phases driven by exceptional points~\cite{Heiss2012,Lee2016,Leykam2017,Xu2017}, the anomalous bulk-boundary correspondence~\cite{Lee2016,Xiong2018,Kunst2018,Yao2018a}, and the non-Hermitian skin effect~\cite{Yao2018b}. Experimental observations of the non-Hermitian topological physics have been reported in classical systems with gain and loss, like the photonic systems~\cite{Xiao2020,Weidemann2020}, the active mechanical metamaterial~\cite{Ghatak2020}, as well as topolectrical circuits~\cite{Helbig2020}, and in quantum simulators, like the nitrogen-vacancy center~\cite{Wu2019,Zhang2020}, where the non-Hermitian effects are engineered by coupling to auxiliary qubits.

As an important source of dissipation and non-Hermiticity, the dynamical noise is ubiquitous and inevitable in the real quantum simulations, especially for the quantum quench dynamics, and can be described by the stochastic Schr\"{o}dinger equation~\cite{Gardiner2004,Gardiner2014}. Without the necessity of applying auxiliary qubits, the quantum simulation using stochastic Schr\"{o}dinger equation may enable a direct and more efficient way to explore non-Hermitian dynamical phases, hence facilitating the discovery of non-Hermitian topological physics with minimal quantum simulation sources. In particular, the controllable noise can provide a fundamental scheme to explore non-Hermitian dissipative quantum dynamics, and the noise effects on the quench-induced dynamical topological phase give rise to rich nonequilibrium topological physics~\cite{Zhang2021noise}. However, the experimental study is currently lacking.

In this article, we report the experimental observation of quench-induced non-Hermitian dynamical topological states by simulating a noising two-dimensional ($2$D) quantum anomalous Hall (QAH) model on an NMR quantum simulator.
Unlike previous experiments using auxiliary qubits~\cite{Wu2019,Zhang2020}, we achieve with advantages the non-Hermitian quench dynamics via simulating the stochastic Schr\"{o}dinger equation and by averaged measurements over different noise configurations~\cite{Gardiner2004,Gardiner2014,Zhang2021noise}. We observe the dynamical topology emerging in the non-Hermitian dissipative quench dynamics on BISs by measuring the time-averaged spin textures in momentum space, and identify two types of dynamical topological transitions classified by distinct dynamical exceptional points by varying the noise strength.
Moreover, the existence of a sweet spot region with the emergent topology being robust under arbitrarily strong noise is experimentally verified. Our experiment demonstrates a feasible technique in simulating dynamical topological physics with minimal sources.

\begin{figure*}
	\includegraphics[scale=0.9]{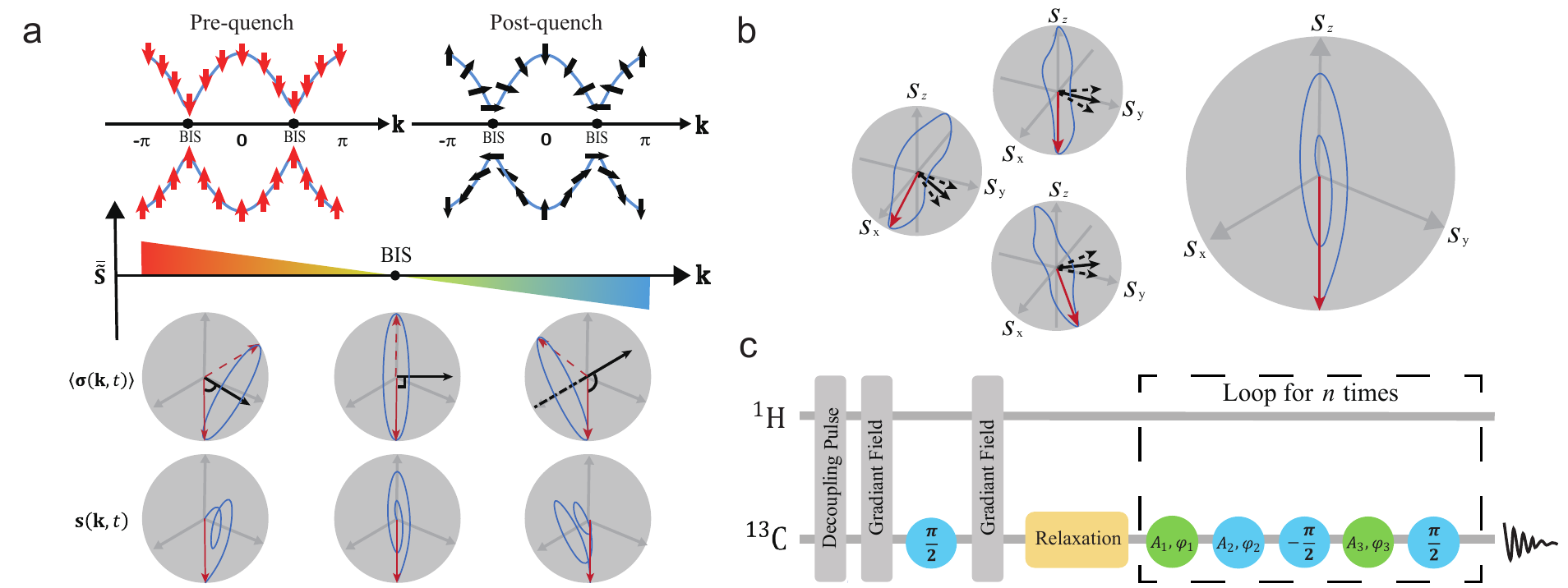}
	\caption{\textbf{Noise-induced non-Hermitian dissipative quantum dynamics and experimental setup.} \textbf{a} Quantum quench process. Upper panel: The system is initialized to the fully polarized ground state $\ket{\downarrow}$ of pre-quench Hamiltonian with $|m_{z}|\gg |\xi_{0}|$. Then $m_{z}$ is quenched to a nontrivial value, and the state evolves under the post-quench Hamiltonian. Middle panel: In the absence of noise, the spin polarization $\langle\boldsymbol{\upsigma}(\mathbf{k},t)\rangle$ (red arrows) precesses with respect to the Hamiltonian vector $\mathbf{h}$ (black arrow). The corresponding trajectory is shown as the blue line. On the BIS where $h_z=0$, the time-averaged spin texture $\overline{\langle \boldsymbol{\upsigma}(\mathbf{k})\rangle}$ vanishes as the Hamiltonian vector $\mathbf{h} = (h_{x},h_{y})$ is orthogonal to the initial state. Across the BIS $\overline{\langle\boldsymbol{\upsigma}(\mathbf{k})\rangle}$ shows nontrivial gradients, which encode the topological invariant. Lower panel: In the presence of noise, the precession axis is distorted, leading to the dissipative dynamics of stochastic averaged spin polarization $\mathbf{s}(\mathbf{k},t)$, as shown by the red arrows and distorted blue trajectories.  \textbf{b} The non-Hermitian dissipative dynamics can be interpreted as the stochastic average over different noise configurations. Here the solid black arrows represent the Hamiltonian vector $\mathbf{h}$ without noise, and the dashed black arrows denote the Hamiltonian vector distorted by time-dependent noise. For each noise configuration (small spheres), the spin polarization is on the surface of Bloch sphere and obeys the unitary dynamics, in spite of the irregular blue trajectory caused by the noise-distorted Hamiltonian vector. However, the average over all noise configurations (big sphere) leads to a globally dissipative effect and a deformation of BISs. \textbf{c} Pulse sequence for simulating the 2D non-Hermitian QAH model. $^1$H is initially decoupled, and $^{13}$C is rapidly prepared to the $\ket{\downarrow}$ state using the nuclear Overhauser effect. The control pulse is designed according to Eq.~\eqref{Trotter}, where the green and blue circles represent rotations about the $x$-axis and $y$-axis with $A$ and $\varphi$ the amplitude and phase, respectively. \label{Figure1}}
\end{figure*}

\section*{RESULTS}
\noindent\textbf{Non-Hermitian QAH model}.
We consider the non-Hermitian 2D QAH model with the magnetic dynamical white noise described by the Hamiltonian
\begin{equation}
\mathcal{H}(\mathbf{k},t)=\mathcal{H}_{\mathrm{QAH}}(\mathbf{k})+\mathbf{w}(\mathbf{k},t)\cdot\boldsymbol{\upsigma},
\label{H_QAH}
\end{equation}
where $\mathcal{H}_{\mathrm{QAH}}(\mathbf{k})=\mathbf{h}(\mathbf{k})\cdot\boldsymbol{\upsigma}$ describes the QAH phase~\cite{Liu2014,Wu2016}, with Bloch vector $\mathbf{h}=(\xi_{\text{so}}\sin k_x,\xi_{\text{so}}\sin k_y,m_z-\xi_0\cos k_x-\xi_0\cos k_y)$. Here $\xi_{0}$ (or $\xi_{\text{so}}$) simulates the spin-conserved (spin-flipped) hopping coefficient, and $m_{z}$ is the magnetic field. The white noise $w_{i}(\mathbf{k},t)$ of strength $\sqrt{w_{i}}$ couples to the Pauli matrix $\sigma_{i}$ and satisfies $\llangle w_i(\mathbf{k},t) \rrangle_{\mathrm{noise}}  =0$ and $\llangle  w_i(\mathbf{k},t)w_j(\mathbf{k},t') \rrangle_{\mathrm{noise}} =w_i\delta_{ij}\delta(t-t')$, where $\llangle \cdot \rrangle_{\mathrm{noise}}$ is the stochastic average over different noise configurations. Without noise, the Hamiltonian $\mathcal{H}_{\mathrm{QAH}}$ hosts nontrivial QAH phase for $0<|m_z|<|\xi_0|$ with Chern number $C_1=\mbox{sgn}(m_z)$, and the phase is trivial for $|m_z|>|\xi_0|$ or $m_z=0$~\cite{Liu2014}. 

The random noise can change the topology of the QAH model, and plays a vital role on the quantum dynamics induced in the present system. We start with the simple situation with a single noise configuration. In this case the quantum dynamics governed by the stochastic Schr\"{o}dinger equation $\mathrm{i}\partial_{t}\vert\psi(\mathbf{k},t)\rangle=\mathcal{H}(\mathbf{k},t)\vert\psi(\mathbf{k},t)\rangle$ describes a random unitary evolution, which can be further converted into the so-called It\^o form~\cite{Gardiner2004,Gardiner2014} in simulation (see Methods for details)
\begin{equation}
	\mathrm{d}\vert\psi(\mathbf{k},t)\rangle=-\mathrm{i}[\mathcal{H}_{{\rm eff}}(\mathbf{k})\mathrm{d}t+\sum_{i}\sqrt{w_{i}}\sigma_{i}\mathrm{d}W_{i}(\mathbf{k},t)]\vert\psi(\mathbf{k},t)\rangle.
\end{equation}
Here $\mathcal{H}_{\mathrm{eff}}=\mathcal{H}_{\mathrm{QAH}}-(\mathrm{i}/2)\sum_{i}w_{i}$ is the effective non-Hermitian Hamiltonian, such that the increment of a Wiener process $W_{i}(\mathbf{k},t)\equiv(1/\sqrt{w_{i}})\int_{0}^{t}\mathrm{d}s\,w_{i}(\mathbf{k},s)$ is independent from the wavefunction function $\vert\psi(t)\rangle$, 
for which we have the It\^o rules $\mathrm{d}t\mathrm{d}W_{i}(t)=0$ and $\mathrm{d}W_{i}(t)\mathrm{d}W_{j}(t)=\delta_{ij}\mathrm{d}t$, and the corresponding expectation value is zero. The formal solution of the above equation reads $\vert\psi(t)\rangle=U(t)\vert\psi(0)\rangle$ with 
\begin{equation}\label{eq:formal solution}
	U(t)=\mathcal{T}\exp\left(-\mathrm{i}\int_{0}^{t}[\mathcal{H}_{{\rm QAH}}\mathrm{d}s+\sum_{i}\sqrt{w_{i}}\sigma_{i}\mathrm{d}W_{i}(s)]\right),
\end{equation}
where $\mathcal{T}$ denotes the time ordering. 
Note that while the equation~\eqref{eq:formal solution} describes a random unitary evolution in the regime with single noise configuration, after the noise configuration averaging the non-Hermitian dissipative quantum dynamics emerges and is captured by the master equation
\begin{equation}\label{eq:non-Hermitian dissipative quantum dynamics}
	\frac{\mathrm{d}\rho(t)}{\mathrm{d}t}=-\mathrm{i}[\mathcal{H}_{{\rm{QAH}}},\rho(t)]+\sum_{i=x,y,z}w_{i}[\sigma_{i}\rho(t)\sigma_{i}-\rho(t)],
\end{equation}
where $\rho(\mathbf{k},t)\equiv\llangle\vert\psi(\mathbf{k},t)\rangle\langle\psi(\mathbf{k},t)\vert\rrangle_{\mathrm{noise}}$ is the stochastic averaged density matrix; see Methods~for details. The configuration averaging is a key point for the present quantum simulation of non-Hermitian dynamical topological phases.

\noindent\textbf{Quantum simulation approach}.
We next develop the quantum simulation approach by introducing discrete Stochastic Schr\"{o}dinger equation for the non-Hermitian dissipative quantum dynamics, since the continuous evolution cannot be directly emulated with digital quantum simulators. Specifically, we discretize the continuous time as $t_{n}=n\tau$ with small time step $\tau$, where the integer $n$ ranges from zero to the total number of time steps $M$. The increment of Wiener process can be simulated by random numbers $\Delta W_{i}(t_{n})=N_{i}(t_{n})\sqrt{\tau}$ for each noise configuration, and we obtain the discretized stochastic Schr\"{o}dinger equation
\begin{equation}
	\vert\psi(\mathbf{k},t_{n+1})\rangle\approx [1-\mathrm{i}\tilde{\mathcal{H}}(\mathbf{k},t_{n})\tau]\vert\psi(\mathbf{k},t_{n})\rangle
\end{equation}
with $\tilde{\mathcal{H}}(\mathbf{k},t_{n})=\mathcal{H}_{\mathrm{eff}}(\mathbf{k})+\sum_{i}\sqrt{w_{i}}\sigma_{i}N_{i}(\mathbf{k},t_{n})/\sqrt{\tau}$.
Here $N_{i}(t_{n})$ is sampled from the standard normal distribution to match the expectation and variance of $\mathrm{d}W_{i}$, and the wavefunction is normalized in each time step. The corresponding unitary evolution operator from time $t_{n}$ to $t_{n+1}$ reads
\begin{equation}\label{eq:evolution operator}
	U(t_{n+1},t_{n})\approx \mathrm{e}^{-\mathrm{i}[\mathcal{H}_{\mathrm{QAH}}+\sum_{i}\sqrt{w_{i}}\sigma_{i}N_{i}(t_{n})/\sqrt{\tau}]\tau},
\end{equation}
leading to the discrete equation of motion
\begin{equation}
	\begin{aligned}
		\rho(t_{n+1}) & \approx \rho(t_{n}) -\mathrm{i}[\mathcal{H}_{\mathrm{QAH}},\rho(t_{n})]\tau \\
		& \qquad{} + \sum_{i}w_{i}[\sigma_{i}\rho(t_{n})\sigma_{i}-\rho(t_{n})]\tau
	\end{aligned}
\end{equation}
in the linear order of $\tau$ after stochastic average, which describes the desired non-Hermitian quantum dynamics. We shall analyse the quality of this discretization versus time step $\tau$ in the experiment. The stochastic average of a physical operator $\mathbf{\hat O}$ at time $t_n$ can now be obtained by
\begin{equation}\label{eq:operatorvalue}
{\mathbf{O}}(\mathbf{k},t_{n})\equiv\llangle\langle\mathbf{\hat O}(\mathbf{k},t_{n})\rangle\rrangle_{\mathrm{noise}}=\mathrm{Tr}[\rho(t_{n})\mathbf{\hat O}].
\end{equation}
This formalism can be directly simulated in experiment.

The above presents the essential idea for simulating the non-Hermitian systems based on the stochastic Schr\"{o}dinger equation. This method is fundamentally different from that applied in the previous experiments~\cite{Wu2019,Zhang2020} using auxiliary qubits, where the non-Hermiticity is obtained from a Hermitian Hamiltonian in the extended Hilbert space by tracing the auxiliary degrees of freedom and careful designs of the quantum circuit with complex unitary operations are required~\cite{Gunther2008, Kawabata2017}. In contrast, our temporal average approach based on the stochastic Schr\"{o}dinger equation saves the resources of qubits and avoids the implementation of complex gates, which benefits the experimental platforms in various scenarios. Moreover, this quantum simulation approach can be directly extended to exploring higher dimensional non-Hermitian topological phases and phase transitions.

\noindent\textbf{Non-Hermitian dynamical topological phases}.\label{sec:Quench dynamics and non-Hermitian dynamical topological phases}
Before presenting the experiment, in this section we briefly introduce the non-Hermitian dynamical topological phases emerging in the quench dynamics described by Eq.~\eqref{eq:non-Hermitian dissipative quantum dynamics} and to be studied in this work.

The system is initially prepared at the fully polarized ground state $\rho_0 = \ket{\downarrow}\bra{\downarrow}$ of a deep trivial Hamiltonian with $|m_z|\gg |\xi_0|$. After quenching $m_z$ to a nontrivial value at time $t_{n}=0$, the system starts to evolve under the post-quench Hamiltonian $\mathcal{H}(\mathbf{k},t_{n})$; see Fig.~\ref{Figure1}a.
Without noise, the spin polarization $\langle \boldsymbol{\upsigma}(\mathbf{k},t_{n})\rangle \equiv\text{Tr}[\boldsymbol{\upsigma}\prod^{n-1}_{i=0}U(t_{n-i},t_{n-i-1})\rho_0\prod^{n-1}_{i=0}U^{\dagger}(t_{i+1},t_{i})]$ precesses with respect to the Hamiltonian vector $\mathbf{h}$; see Fig.~\ref{Figure1}a. The post-quench QAH phase can be determined by the dynamical topology emerging on BISs~\cite{Zhang2018}, identified as the momentum subspaces with $h_{z}=0$, where the initial state is perpendicular to the SO field $\mathbf{h}_{{\rm{so}}}\equiv(h_{x},h_{y})$, leading to vanishing time-averaged spin polarizations.

In the presence of non-Hermiticity, the precession axis for each noise configuration is distorted, leading to the deformation for the BISs and dissipative effect. To characterize the noise effect, the spin polarization needs to be stochastically averaged as
\begin{equation}\label{eq:spinoperatorvalue1}
\mathbf{s}(\mathbf{k},t_{n})\equiv\llangle\langle\boldsymbol{\upsigma}(\mathbf{k},t_{n})\rangle\rrangle_{\mathrm{noise}}=\mathrm{Tr}[\rho(t_{n})\boldsymbol{\upsigma}]
\end{equation}
over different noise configurations [see Fig.~\ref{Figure1}b]. Compared to the spin polarization $\langle \boldsymbol{\upsigma}(\mathbf{k},t_{n})\rangle$ without noise, the stochastic averaged $\mathbf{s}(\mathbf{k},t_{n})$ follows the non-Hermitian dynamics and exhibits dephasing and amplitude decaying effects.
We compensate the amplitude decay by rescaling $\mathbf{s}(\mathbf{k},t_{n})$, leading to the rescaled spin polarization $\tilde{\mathbf{s}}(\mathbf{k},t_{n})\equiv\mathbf{s}_{0}(\mathbf{k})+\mathbf{s}_{+}(\mathbf{k})\mathrm{e}^{-\mathrm{i}\omega(\mathbf{k})t_{n}}+\mathbf{s}_{-}(\mathbf{k})\mathrm{e}^{+\mathrm{i}\omega(\mathbf{k})t_{n}}$, where the coefficients $\mathbf{s}_{0,\pm}$ and oscillation frequency $\omega$ are extracted from the experimental data by fitting; see Methods. Similar to the noiseless case, the time average
\begin{equation}\label{eq:spinoperatorvalue1}
\overline{\tilde{\mathbf{s}}(\mathbf{k})}\equiv\frac{1}{M}\sum^{M-1}_{n=0}\tilde{\mathbf{s}}(\mathbf{k},t_{n})
\end{equation}
vanishes on the deformed BISs (dubbed as dBISs)~\cite{Zhang2021noise}, with the number of steps $M$ being large enough to minimize the error. The non-Hermitian dynamical topological phase is captured by the dynamical invariant $\mathcal{W}\equiv\frac{1}{2\uppi}\oint_{\rm{dBIS}}\mathbf{g}(\mathbf{k})\mathrm{d}\mathbf{g}(\mathbf{k})$, which describes the winding of dynamical field $\mathbf{g}(\mathbf{k})=(1/\mathcal{N}_{\mathbf{k}})\partial_{k_{\perp}}(\overline{\tilde{s}_{x}(\mathbf{k})},\overline{\tilde{s}_{y}(\mathbf{k})})$ on the dBISs. Here $k_{\perp}$ is perpendicular to the dBISs and $\mathcal{N}_{\mathbf{k}}$ is a normalization factor. Under the dynamical noise, the non-Hermitian dynamical topological phases and phase transitions may be induced, as studied in the experiment presented below.

\begin{figure}
	\centering
	\includegraphics[scale=0.4]{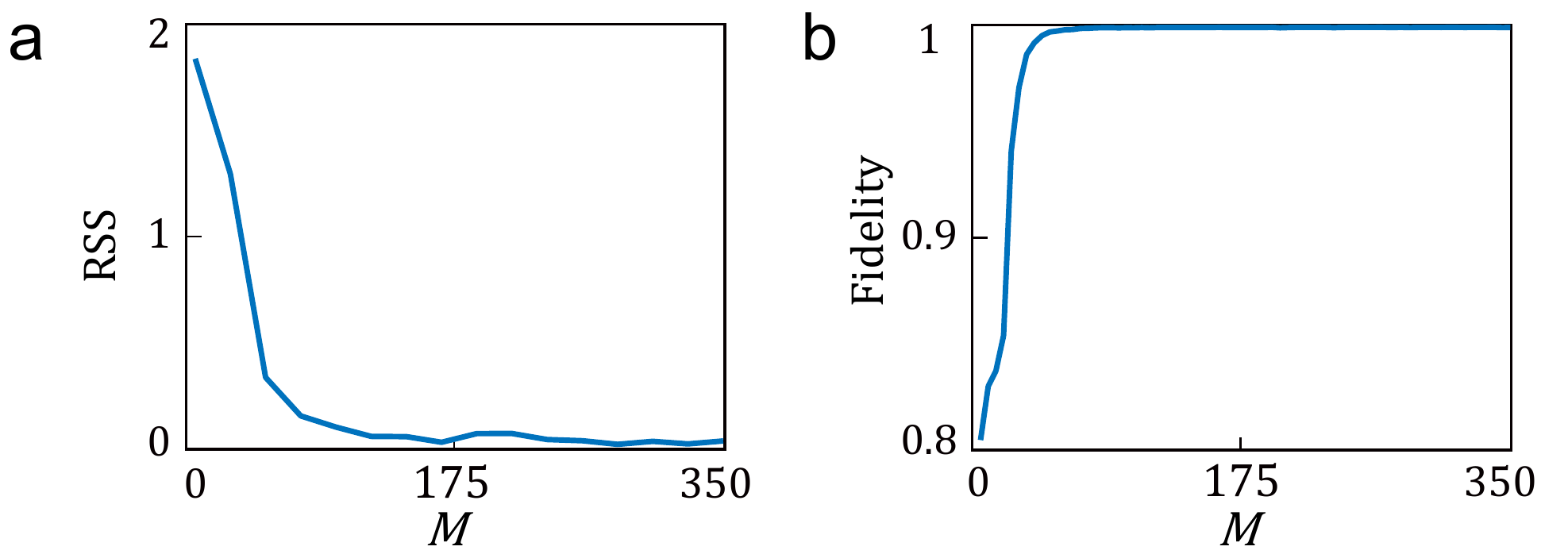}
	\caption{\textbf{Quality of discrete Stochastic Schr{\"o}dinger equation approach versus the total number of time steps.} \textbf{a} Residual sum of squares (RSS) of the stochastic averaged spin polarizations $\mathbf{s}$ at momentum $\mathbf{k}=(1.2857, -1.8)$ between the discrete evolution and continuous evolution at time $t=30$ ms for different numbers of discrete time steps $M$. The stochastic average is performed over 5,000 noise configurations. \textbf{b} Corresponding average fidelity for different numbers of discrete time steps $M$. When the number of time steps is greater than $100$, the fidelity is over $0.99$. Here we set $w_x=0.05\xi_0$, $w_y=0$, $w_z=0.01\xi_0$, and $\xi_{\text{so}}=0.2\xi_0$.\label{Figure2} }
\end{figure}

\noindent\textbf{Experimental setup}.\label{sec:Experimental setup}
\begin{figure}
	\centering
	\includegraphics[scale=0.5]{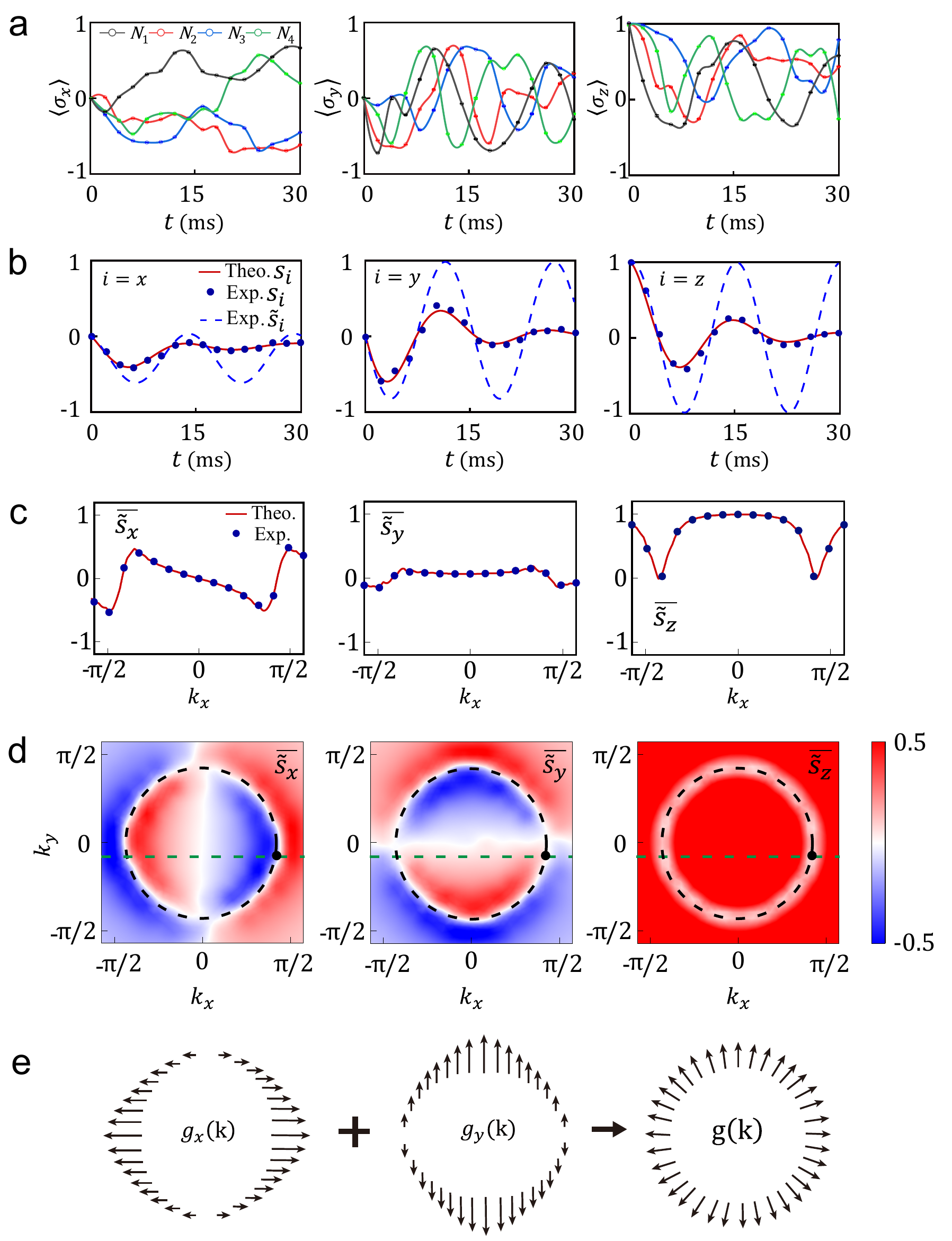}
	\caption{\textbf{Weak noise regime of the non-Hermitian QAH model.}  \textbf{a} Spin polarization $\langle \boldsymbol{\upsigma}(\mathbf{k},t)\rangle$ under four different white noise configurations at momentum $\mathbf{k}= (1.286, -0.257)$. No notable decay is observed for a single type of noise. \textbf{b} Corresponding stochastic averaged spin polarization $\mathbf{s}(\mathbf{k},t)$ over all noise configurations and the rescaled spin polarization $\tilde{\mathbf{s}}(\mathbf{k},t)$, which has the minimum oscillation frequency $\omega$ on the dBIS at momentum $\mathbf{k}= (1.286, -0.257)$. The system is dissipative after stochastic average as an outcome of non-Hermitian dynamics. \textbf{c} Time-averaged rescaled spin polarizations $\overline{\tilde{s}_{i}(\mathbf{k})}$ for fixed $k_y = -0.257$ and $k_x \in [-1.8, 1.8]$. The zero values represent dBIS points. \textbf{d} Time-averaged spin texture by discretizing the momentum space $k_x, k_y \in [-1.8, 1.8]$ into a $15\times 15$ lattice, where the dBIS momenta can be outlined according to the zero values. The black dot and green dashed line correspond to the case of (a, b) and \textbf{c}, respectively. \textbf{e} Dynamical field $\mathbf{g}=(g_{x},g_{y})$ (black arrows) obtained from the time-averaged spin texture. Here we set $w_x=0.05\xi_0$, $w_y=0$, $w_z=0.01\xi_0$, and $\xi_{\text{so}}=0.2\xi_0$.
\label{Figure3}}
\end{figure}
The demonstration is performed on the NMR quantum simulator. The sample is the $^{13}$C-labeled chloroform dissolved in acetone-d6, with $^{13}$C and $^{1}$H nuclei denoted as two qubits. The 2D QAH model is simulated by the qubit $^{13}$C, while the other qubit $^{1}$H enhances the signal by Overhauser effect [see Fig.~\ref{Figure1}c and Methods]. In the double-rotating frame, the total Hamiltonian of this sample is
\begin{equation}
\mathcal{H}_{e}=\mathrm{\uppi} J\sigma_z^1\sigma_z^2/2+\sum_{i=1}^2 \mathrm{\uppi} B_i\left(\cos \phi_i\sigma_x^i+\sin \phi_i\sigma_y^i\right),
\label{Hnmr}
\end{equation}
where $J=215$ Hz is the coupling strength, $B_i$ is the amplitude of the control pulse, and $\phi_i$ is the phase. We firstly initialize the system into the fully polarized state $\ket{\downarrow}$ using the nuclear Overhauser effect~\cite{Noggle1971}. Then we quench $m_z$ to the nontrivial region with $|m_z|<2\xi_0$ and allow the system to evolve under the effective Hamiltonian $\tilde{\mathcal{H}}$, in which the non-Hermitian constant term $\mathrm{i}\sum_{i}w_{i}$ can be ignored. The evolution is realized by the Trotter approximation combined with control pulse optimizations as follows.

We study the non-Hermitian dissipative quantum dynamics from time $t=0$~ms to $30$~ms. For each noise configuration, numerical results show that the discrete evolution approximates the continuous evolution of the stochastic Schr\"{o}dinger equation quite well, when the total number of time steps is greater than $100$; see Fig.~\ref{Figure2}. In experiment, we discretize the time into  $300$ segments, such that the Hamiltonian in each interval is approximately time-independent. As the interval $\tau$ is sufficiently small, the evolution in the $n$-th step can be realized using the first-order Trotter decomposition:
\begin{equation}
U(t_{n+1},t_{n})\approx \mathrm{e}^{-\mathrm{i}\eta_x\sigma_x\tau}\mathrm{e}^{-\mathrm{i}\eta_y\sigma_y\tau}\mathrm{e}^{-\mathrm{i}\eta_z\sigma_z\tau}
\label{Trotter}
\end{equation}
with $\eta_{x, y}=\xi_{\text{so}}\sin  k_{x, y}+\sqrt{w_{x, y}} N_{x, y}(t_n)/\sqrt{\tau}$ and $\eta_z=m_z-\xi_0\cos  k_x-\xi_0\cos  k_y+\sqrt{w_z}N_{z}(t_n)/\sqrt{\tau}$. Here $\xi_0$ is set to $1$~kHz. Each term on the right-hand side represents a single-qubit rotation with rotating angle $2\eta_i\tau$ along axis $\sigma_i$, which can be experimentally realized by tuning the amplitude and phase of the control pulse in Eq.~\eqref{Hnmr} ($z$-rotation can be indirectly realized via $x$- and $y$-rotations), with further pulse optimization techniques to reduce control errors; see Fig.~\ref{Figure1}c.

We measure the spin polarization $\langle\boldsymbol{\upsigma}(\mathbf{k},t)\rangle$ for single noise configuration at every $20\tau$ interval. After averaging over all noise configurations, we obtain the stochastic averaged spin polarization $\mathbf{s}(\mathbf{k},t)$, from which the rescaled spin polarization $\tilde{\mathbf{s}}(\mathbf{k},t)$ can be constructed by fitting. We repeat the above procedures for the whole momentum space to obtain the time-averaged spin textures $\overline{\tilde{\mathbf{s}}(\mathbf{k})}$.

\noindent\textbf{Experimental results}.\label{sec:Results}
We start from the weak noise regime, where the noise strength is chosen as $w_x=0.05\xi_0$, $w_y=0$, and $w_z=0.01\xi_0$ with $\xi_{\mathrm{so}}=0.2\xi_{0}$. The system is quenched to the topological phase with $m_z = 1.2\xi_0$.
In Fig.~\ref{Figure3}a, we plot the spin polarization $\langle\boldsymbol{\upsigma}(t)\rangle$ at the momentum $\mathbf{k}=(1.286, -0.257)$ for four different noise configurations. For each noise configuration, no notable decay exists in the spin polarization, manifesting the unitary evolution. However, after averaged over all noise configurations, the system clearly exhibits the non-Hermitian dissipative quantum dynamics; see Fig.~\ref{Figure3}b.

\begin{figure}
	\centering
	\includegraphics[scale=0.43]{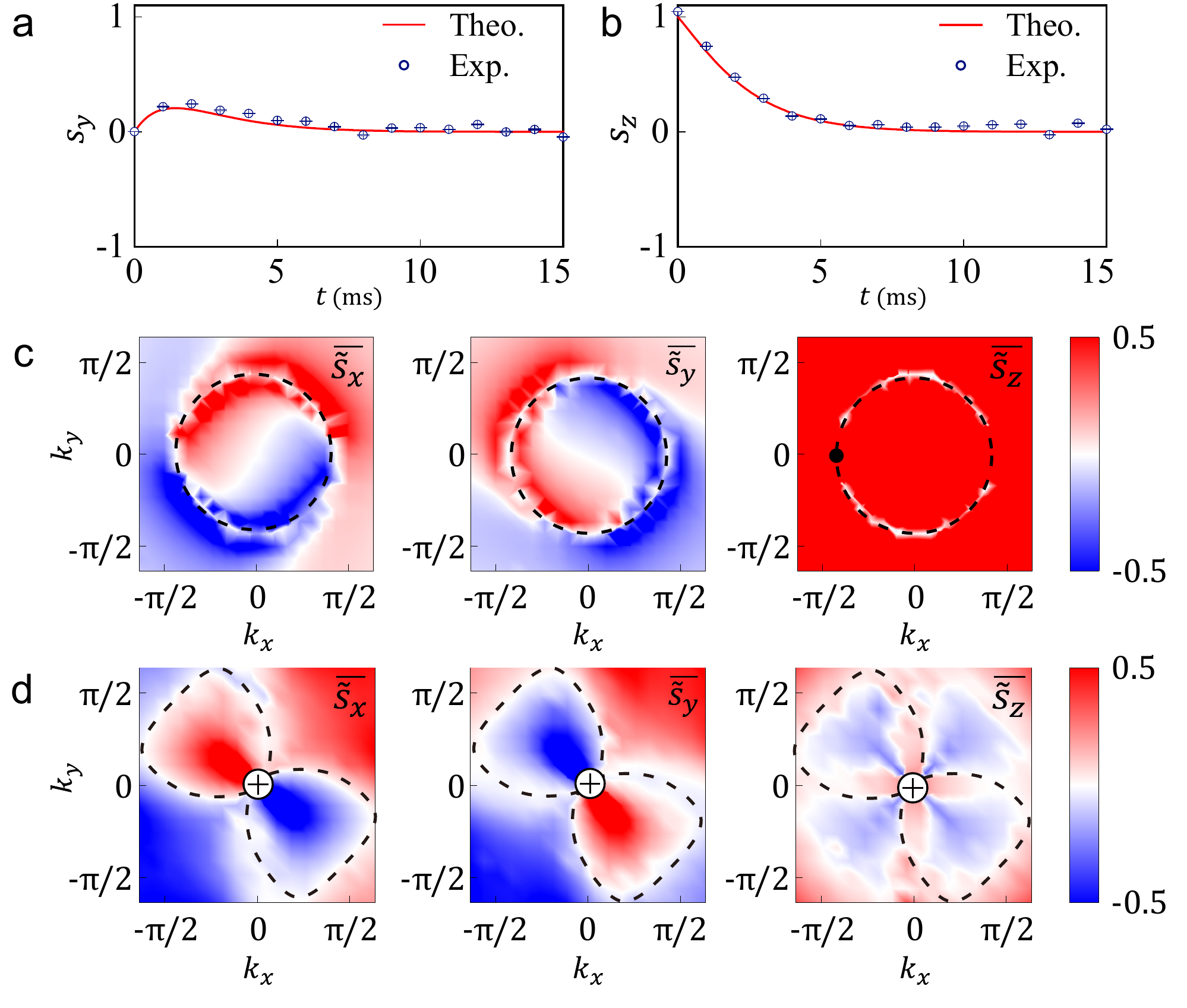}
	\caption{\textbf{Strong noise regime of the non-Hermitian QAH model.} \textbf{a} For $w_x=0.1\xi_0$, $w_y=0.05\xi_0$, $w_z=0.45\xi_0$ and $\xi_{\text{so}}=0.2\xi_0$, the stochastically averaged spin polarization $s_{y}(\mathbf{k},t)$ presents no oscillation.  \textbf{b} At the same noise level, $s_{z}(\mathbf{k},t)$ decays to $0$ without oscillation. \textbf{c} Time-averaged rescaled spin textures $\overline{\tilde{s}_{i}(\mathbf{k})}$, with $i=x,y,z$. Other than small deformation, singularities (black dot) emerge on the dBIS momenta (type-I dynamical transition). \textbf{d} $\overline{\tilde{s}_{i}(\mathbf{k})}$ under stronger noise of $w_x=1.6\xi_0$, $w_y=0$, $w_z=0.8\xi_0$, and $\xi_{\text{so}}=2\xi_0$. The dBIS deforms drastically and connects to the topological charge at $\mathbf{k}=0$ (type-II dynamical transition). \label{Figure4}}
\end{figure}

\begin{figure}
	\centering
	\includegraphics[scale=0.4]{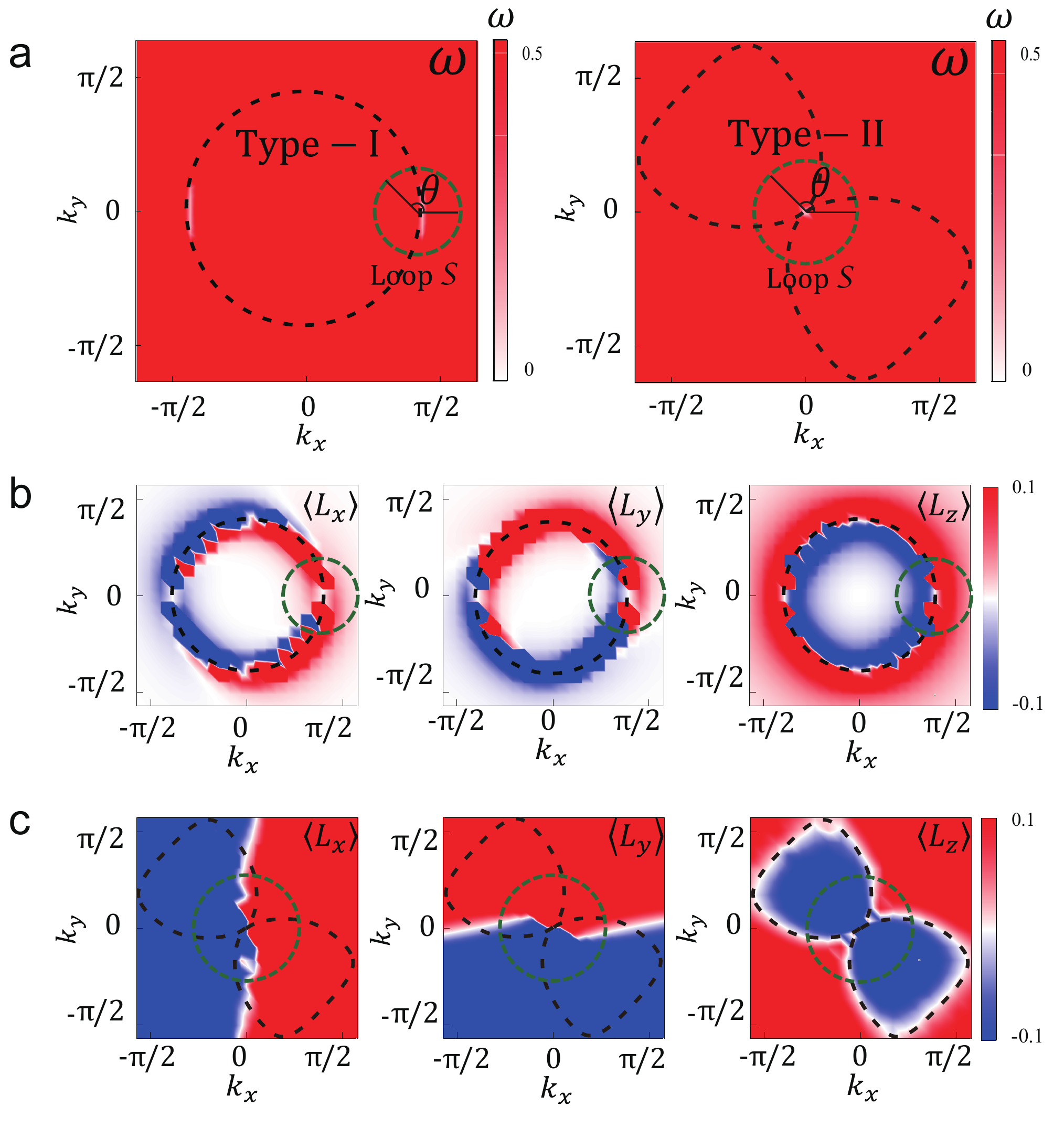}
	\caption{\textbf{Exceptional points and Liouvillian polarizations for dynamical transitions.} \textbf{a} Experimental oscillation frequency for the type-I and type-II dynamical transition shown in Fig.~\ref{Figure4}. The exceptional points are captured by momenta with vanishing $\omega$ and are enclosed by the loop $\mathcal{S}$ (green dashed circles) of the form $(x_{0}+r\cos{\theta}, y_{0}+r\sin{\theta})$ for the convenience of view [see Fig.~\ref{Figure10} in Methods for more detailed area of exceptional points], which connect with the dBIS (black dashed lines) in both types of dynamical transitions. 
Particularly, for the type-II transition, the exceptional point locates at the charge momentum $\mathbf{k}=0$, to which the dBIS is deformed. \textbf{b, c} Measured Liouvillian polarization $\langle \mathbf{L} \rangle$ for the type-I and type-II dynamical transition, respectively.\label{Figure5}}
\end{figure}

\begin{figure*}
	\centering
	\includegraphics[scale=0.43]{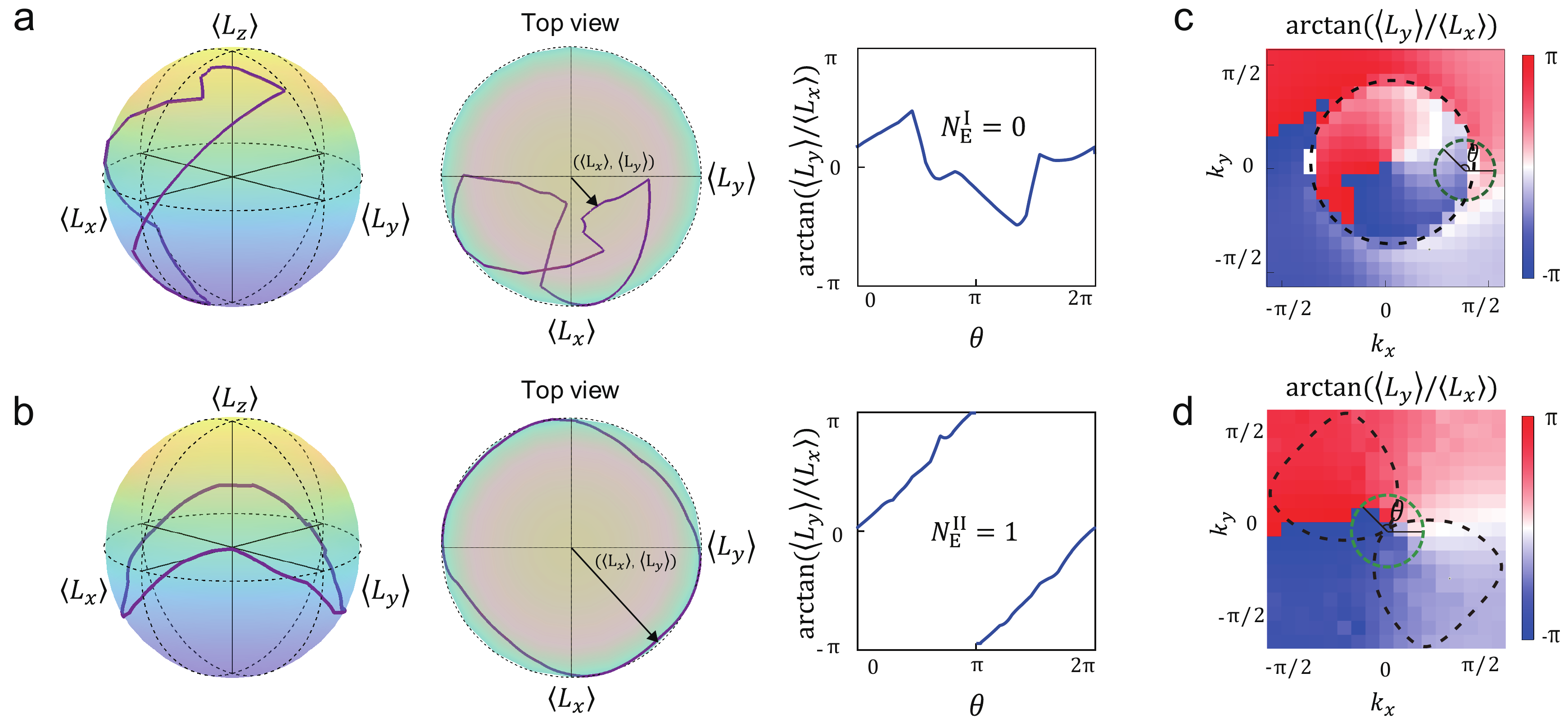}
	\caption{\textbf{Winding number $N_{\mathrm{E}}$ for dynamical transitions.} \textbf{a} Trajectory of Liouvillian polarization $\langle\mathbf{L}\rangle$ along the loop $\mathcal{S}$ (green dashed circle in \textbf{c}) for type-I dynamical transition. From top view along the $\langle L_{z}\rangle$ axis, the trajectory does not encircle the origin point of $\langle L_{x}\rangle$-$\langle L_{y}\rangle$ plane but trivially returns to its initial position when $\theta$ changes from $0$ to $\pi$, manifesting a trivial winding number $N_{\mathrm{E}}^{\text{I}}=0$. \textbf{b} Trajectory of Liouvillian polarization along the loop $\mathcal{S}$ (green dashed circle in \textbf{d}) for type-II dynamical transition. Unlike type-I transition, the trajectory projected on the $\langle L_{x}\rangle$-$\langle L_{y}\rangle$ plane now encircles the origin point. Particularly, there is a jump in the values of $\mathrm{arctan}\langle L_y\rangle/\langle L_x\rangle$ along the loop $\mathcal{S}$, manifesting a nontrivial winding number $N_{\mathrm{E}}^{\text{II}}=1$. \textbf{c, d} Distribution of $\mathrm{arctan}\langle L_y\rangle/\langle L_x\rangle$ in the momentum space for type-I (\textbf{c}) and type-II (\textbf{d}) dynamical transitions. As long as the loop $\mathcal{S}$ does not cross any other singular points, the winding number $N_{\mathrm{E}}$ of exceptional point remains unchanged. Especially, for exceptional points not in contact with topological charges, $\mathrm{arctan}\langle L_y\rangle/\langle L_x\rangle$ changes continuously. On the other hand, the charge momentum $\mathbf{k}=0$ is always singular.\label{Figure6}}
\end{figure*}

Fig.~\ref{Figure3}c shows the measured time-averaged spin textures $\overline{\tilde{s}_{i}(\mathbf{k})}$ with fixed $k_y = -0.257$ and $k_x \in [-1.8, 1.8]$, obtained by rescaling the stochastic averaged spin polarization $\mathbf{s}(\mathbf{k},t)$. The momenta with vanishing values represent dBIS points. To obtain the 2D time-averaged spin texture, we discretize the whole momentum space $k_x, k_y \in [-1.8, 1.8]$ into a $15\times 15$ lattice and repeat the above measurements. The results are shown in Fig.~\ref{Figure3}d, from which the dBIS momenta can be identified. Although the corresponding shape is slightly deformed from the ideal BIS with $h_{z}=0$ in the absence of noise (see Methods), it is obvious that under weak noise, the dynamical field $\mathbf{g}(\mathbf{k})$ can be defined everywhere on dBIS and characterizes the nontrivial non-Hermitian dynamical topological phase [see Fig.~\ref{Figure3}e]. Indeed, this emergent dynamical topology is robust against the weak noise and is protected by the finite minimal oscillation frequency on the dBISs, serving as a bulk gap for the dynamical topological phase. The experimental minimum oscillation frequency on dBISs is given by $\omega_{\rm min}=0.4175$~kHz, close to the theoretical value $0.4063$~kHz [Fig.~\ref{Figure3}b]. Further, this non-Hermitian dynamical topological phase may break down under strong noise, with two types of dynamical transition being observed below. 

We now increase the noise strength to a strong regime with $w_x=0.1\xi_0$, $w_y=0.05\xi_0$, and $w_z=0.45\xi_0$. The averaged spin polarization is measured in the same way as in the weak noise regime. However, the quench dynamics are essentially different, where the spin polarization $\mathbf{s}(t)$ at certain momenta, for instance $k_x= -1.286$ and $k_y = -0.257$, displays pure decay without oscillation; see Figs.~\ref{Figure4}a and \ref{Figure4}b.
For these momenta, the dynamical field $\mathbf{g}$ vanishes.
In Fig.~\ref{Figure4}c, we show the corresponding spin textures. From the result for $\overline{\tilde{s}_{z}}$, we find that singularities emerges on the dBISs and interrupt their continuity. Thus the dBIS breaks down, while the deformation of the shape of dBIS is small, and the non-Hermitian dynamical topological phase transition occurs.
In Fig.~\ref{Figure4}d, we increase the noise strength to $w_x=1.6\xi_0$, $w_y=0$, $w_z=0.8\xi_0$ and set a strong SO coupling coefficient with $\xi_{\text{so}}=2\xi_0$. A qualitatively different dynamical transition is uncovered, where the dBISs are dramatically deformed by the noise and are connected to the topological charge at $\mathbf{k}=0$. Due to this singularity, the dynamical topology also breaks down. The above two qualitatively different phenomena are referred to as type-I and type-II dynamical transitions, respectively, which we examine below in more detail.

We notice that the equilibrium topological phase transition usually corresponds to the close of energy gap. In the nonequilibrium regime, the analogous quantity is the oscillation frequency. Here we observe the corresponding momentum distribution in Fig.~\ref{Figure5}a. One can see that the oscillation frequency is in general nonzero but may vanish on certain dBISs momenta when these two types of dynamical transition occur, i.e. $\omega_{\rm min}(\bold k_c)\rightarrow0$. Indeed, the momenta ($\bold k_c$) with just vanishing oscillation frequency are exceptional points of the Liouvillian superoperator, on which the eigenvectors $\mathbf{s}_{\pm}^{L(R)}$ coalesce~\cite{Zhang2021noise}. Thus the dynamical transitions are driven by exceptional points with vanishing oscillation frequency on dBISs.
To further distinguish these two types of dynamical transition and the corresponding exceptional points, we treat the Liouvillian superoperator as a three-level system; see Methods. The coefficient $\mathbf{s}_{+}$ of rescaled dynamical spin polarization $\tilde{\mathbf{s}}(\mathbf{k},t)$ contains the information of corresponding eigenvectors $\mathbf{s}_{\pm}^{L(R)}$. Like the spin-1 system, we measure the Liouvillian polarization $\langle L_{\alpha}\rangle\equiv \mathbf{s}_{+}^{\dagger}L_{\alpha}\mathbf{s}_{+}$ to characterize the Liouvillian superoperator. Here the operator $L_{\alpha}$ is defined as
\begin{equation}
	L_{x}=\begin{pmatrix}0 & 0& 0\\0 & 0 & -\mathrm{i}\\0 & \mathrm{i} & 0\end{pmatrix},\qquad{}L_{y}=\begin{pmatrix}0&0&\mathrm{i}\\0&0&0\\-\mathrm{i}&0&0
	\end{pmatrix},
\end{equation}
and $L_{z}=\mathrm{i}[L_y,L_x]$, which satisfies $[L_{\alpha},L_{\beta}]=\mathrm{i}\epsilon^{\alpha\beta\gamma}L_{\gamma}$. The measured momentum distribution of these quantities in experiment is shown in Figs.~\ref{Figure5}b and \ref{Figure5}c, from which an important feature of exceptional points is observed that the component $\langle L_{x}\rangle\approx0$ and $\langle L_{y}\rangle\approx0$ vanish on these points while $\langle L_{z}\rangle$ is in general nonzero [e.g., see Fig.~\ref{Figure5}c]. Therefore, the exceptional points are actually the singularities in the two-component vector field $(\langle L_{x}\rangle,\langle L_{y}\rangle)$.

With this observation and to characterize the exceptional points, we consider the Liouvillian polarization on a small loop $\mathcal{S}$ enclosing the exceptional points, as shown in Figs.~\ref{Figure6}a and \ref{Figure6}b. Although the component $\langle L_{z}\rangle$ is nonzero on this loop, the trajectory projected on the $\langle L_{x}\rangle$-$\langle L_{y}\rangle$ plane indeed defines a winding number~\cite{Zhang2021noise}
\begin{equation}
	N_{\mathrm{E}}=\frac{1}{2\mathrm{\uppi}}\oint_\mathcal{S}\mathrm{d}(\mathrm{arctan}\langle L_y\rangle/\langle L_x\rangle),
\end{equation}
which distinguishes the two types of dynamical transitions. We observe that for type-I transition, the winding number $N_{\mathrm{E}}=0$ is trivial, while the winding $N_{\mathrm{E}}=1$ is nontrivial for the type-II dynamical transition. Consequently, these distinct exceptional points on dBISs shows the fundamental difference between the type-I and type-II dynamical transitions. Moreover, regardless of the shape and size of the loop $\mathcal{S}$, the winding number $N_{\mathrm{E}}$ only depends on the topological properties of the enclosed exceptional points as long as the loop does not cross any other singular points; see Figs.~\ref{Figure6}a and \ref{Figure6}b. Here we note that the topological charges are always singularities of the field $(\langle L_{x}\rangle,\langle L_{y}\rangle)$ and have nontrivial winding number~\cite{Zhang2021noise} [see Figs.~\ref{Figure6}c and \ref{Figure6}d]. The loop $\mathcal{S}$ should be introduced without enclosing any non-exceptional charge momentum in characterizing the dynamical transitions and corresponding exceptional points. This also tells that the type-II dynamical transition is similar to the equilibrium topological phase transition, in which the topological charges serve as singular points and the transition occurs when they pass through the BISs~\cite{Zhang2018,Zhang2019}. On the other hand, the type-I transition is a peculiar feature of the quench-induced non-Hermitian dynamical topological phase transition.

\begin{figure}
	\centering
	\includegraphics[scale=0.5]{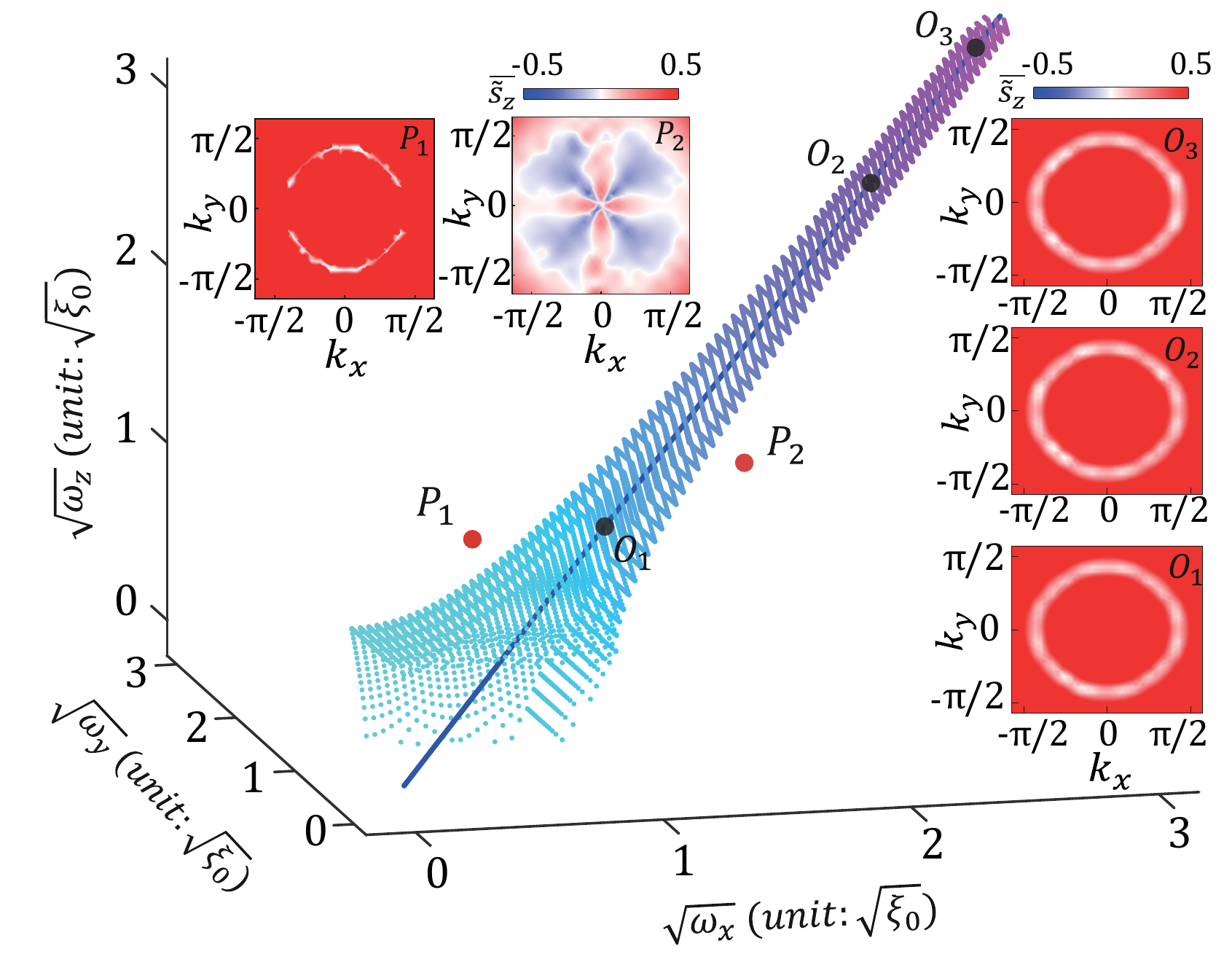}
	\caption{\textbf{Sweet spot region in the non-Hermitian QAH model.} The critical surface of the region is coordinated by color points, where a particular example is the straight line with $w_x=w_y=w_z$. In experiment, we increase the noise strength in each direction to $10\xi_0$, and measure the time-averaged rescaled spin texture $\overline{\tilde{s}_{z}(\mathbf{k})}$. The three cases $O_1$, $O_2$, and $O_3$ in the sweet spot region with noise intensities $0.5\xi_0$, $5\xi_0$ and $10\xi_0$ are plotted. For comparison,  two cases $P_1$ and $P_2$ outside the region are shown as well. The dBIS is always stable and no singularities are observed in the sweet spot region. \label{Figure7}}
\end{figure}

Although the non-Hermitian dynamical topological phase may typically be destroyed in the strong noise regime, a quite interesting feature of the present system is the existence of a sweet spot region satisfying~\cite{Zhang2021noise}
\begin{align}
\max[(w_y-w_x)\xi^2_0/\xi^2_{\text{so}}-2|\xi_{\text{so}}|]<w_z-w_x\nonumber\\
<\min [(w_y-w_x)\xi^2_0/\xi^2_{\text{so}}+2|\xi_{\text{so}}|],
\end{align}
in which regime the dynamical topology is always robust at any finite noise strength, as characterized by the taper-type region in Fig.~\ref{Figure7}.
In particular, for the central line with $w_x=w_y=w_z$, we experimentally increase the noise strength $w_{i}$ in each direction from $0.5\xi_0$ to a very large value $w_{i}\simeq10\xi_0$ (points $O_{1,2,3}$) and measure the corresponding time-averaged dynamical spin textures. We observe that although the noise strength is much large compared with all other energy scales, the dBIS in $\overline{\tilde{s}_{z}}$ remains stable, without suffering singularities. Inside the taper-type region the dynamical topology is well-defined on the dBIS, in sharp contrast to outside points ($P_{1,2}$). The experimental confirmation of this sweet spot region may offer guidance in designing noise-tolerant topological devices.

\section*{DISCUSSION}\label{sec:Conclusion}

We have experimentally reported the quantum simulation of non-Hermitian quantum dynamics for a 2D QAH model coupled to dynamical noise based on a stochastic average approach of the stochastic Schr\"{o}dinger equation, and simulated non-Hermitian dynamical topological phases and phase transitions. Our method does not require the ancillary qubits and careful designs of complex unitary gates, hence saving the simulation sources and
avoiding the implementation of complex gates in experiment. The dynamical topological physics driven by dynamical noise has been observed, including the stability of non-Hermitian dynamical topological states protected by the minimal oscillation frequency of quench dynamics under weak noise and two basic types of dynamical topological transitions driven by strong noise and classified by distinct exceptional points. Moreover, a sweet spot region is observed, where the non-Hermitian dynamical topological phase survives at arbitrarily strong noise.

Our experiment has shown an advantageous quantum simulation approach to explore the non-Hermitian dynamical topological physics, in which only minimal number of qubits are used. This approach is directly applicable to high dimensions by taking into account more, but still minimal number of qubits, in which the rich phenomena are expected, and also to other digital quantum simulators.

\section*{METHODS}

\noindent\textbf{Stratonovich stochastic Schr\"odinger equation}.
We consider the non-Hermitian 2D QAH model \eqref{H_QAH} with the magnetic dynamical white noise $w_{i}(\mathbf{k},t)$. Since the dynamical white
noise is in some sense infinite, the dynamical equation $\partial_{t}\vert\psi(\mathbf{k},t)\rangle=-\mathrm{i}\mathcal{H}(\mathbf{k},t)\vert\psi(\mathbf{k},t)\rangle$
cannot be considered as an ordinary differential equation. Instead, it should
be regarded as an integral equation
\begin{align}
&\vert\psi(\mathbf{k},T)\rangle-\vert\psi(\mathbf{k},0)\rangle
=\int^{T}_{0}\mathrm{d}\vert\psi(\mathbf{k},t)\rangle \nonumber\\
&=-\mathrm{i}\int^{T}_{0}[\mathcal{H}_{{\rm{QAH}}}(\mathbf{k})\mathrm{d}t+\sum_{i}\sqrt{w_{i}}\sigma_{i}\mathrm{d}W_{i}(\mathbf{k},t)]\vert\psi(\mathbf{k},t)\rangle,
\end{align}
where $W_{i}(\mathbf{k},t)=(1/\sqrt{w_{i}})\int_{0}^{t}\mathrm{d}s\,w_{i}(\mathbf{k},s)$
is a Wiener process.  For brevity, the symbols of integration are
usually dropped, leading to the stochastic Schr\"odinger equation
\begin{equation}\label{eq:Stratonovich stochastic Schrodinger equation}
\mathrm{d}\vert\psi(\mathbf{k},t)\rangle=-\mathrm{i}[\mathcal{H}_{{\rm{QAH}}}(\mathbf{k})\mathrm{d}t+\sum_{i}\sqrt{w_{i}}\sigma_{i}\mathrm{d}W_{i}(\mathbf{k},t)]\vert\psi(\mathbf{k},t)\rangle.
\end{equation}

In general, there are two definitions of stochastic integration, i.e. the Stratonovich form
\begin{equation}
(\mathbf{S})f(t)\mathrm{d}W(t)\equiv\frac{1}{2}[f(t+\mathrm{d}t)+f(t)][W(t+\mathrm{d}t)-W(t)]
\end{equation}
and the It\^o form
\begin{equation}
(\mathbf{I})f(t)\mathrm{d}W(t)\equiv f(t)[W(t+\mathrm{d}t)-W(t)].
\end{equation}
The basic difference is that the integrand $f(t)$ and the increment
$\mathrm{d}W(t)$ are independent of each other in the It\^o form,
namely $\llangle f(t)\mathrm{d}W(t)\rrangle_{{\rm{noise}}}=f(t)\llangle\mathrm{d}W(t)\rrangle_{{\rm{noise}}}=0$,
while they are not independent in the Stratonovich form. The Schr\"odinger
equation~\eqref{eq:Stratonovich stochastic Schrodinger equation} must be interpreted as a Stratonovich stochastic differential
equation~\cite{Gardiner2004,Gardiner2014}, such that the quantum mechanical probability is preserved, i.e. $\mathrm{d}\langle\psi(t)\vert\psi(t)\rangle=0$. 

\noindent\textbf{Converting into the It\^o form}.
\label{sec:Converting into the Ito form}
Since the wavefunction $\vert\psi(t)\rangle$ and the increment $\mathrm{d}W_{i}(t)$ are not independent in the Stratonovich form, it is usually convenient to convert the Stratonovich stochastic Schr\"odinger equation~\eqref{eq:Stratonovich stochastic Schrodinger equation} into the It\^o form, which takes the form
\begin{equation}\label{eq:Ito stochastic Schrodinger equation}
(\mathbf{I})\mathrm{d}\vert\psi(t)\rangle=-\mathrm{i}[\mathcal{H}_{{\rm eff}}\mathrm{d}t+\sum_{i}\alpha_{i}\mathrm{d}W_{i}(t)]\vert\psi(t)\rangle.
\end{equation}
Due to $(1/2)(\vert\psi(t+\mathrm{d}t)\rangle+\vert\psi(t)\rangle)=[\mathbf{1} -(\mathrm{i}/2)(\mathcal{H}_{{\rm eff}}\mathrm{d}t+\sum_{i}\alpha_{i}\mathrm{d}W_{i}(t))]\vert\psi(t)\rangle$, we have the following relation between the Stratonovich integral and the It\^o integral
\begin{equation}
 (\mathbf{S})\mathrm{d}W_{i}(t)\vert\psi(t)\rangle= (\mathbf{I})\mathrm{d}W_{i}(t)\vert\psi(t)\rangle-(\mathrm{i}/2)\alpha_{i}\mathrm{d}t\vert\psi(t)\rangle,
\end{equation}
where we have used the It\^o rules $\mathrm{d}t\mathrm{d}W_{i}(t)=0$
and $\mathrm{d}W_{i}(t)\mathrm{d}W_{j}(t)=\delta_{ij}\mathrm{d}t$
for the increment of a Wiener process. Substituting this into the It\^o stochastic Schr\"odinger equation~\eqref{eq:Ito stochastic Schrodinger equation}, we obtain
\begin{equation}
	(\mathbf{S})\mathrm{d}\vert\psi(t)\rangle = -\mathrm{i}[(\mathcal{H}_{{\rm eff}}+\frac{\mathrm{i}}{2}\sum_{i}\alpha_{i}^{2})\mathrm{d}t+\sum_{i}\alpha_{i}\mathrm{d}W_{i}(t)]\vert\psi(t)\rangle.
\end{equation}
Compared with the original Stratonovich stochastic Schr\"odinger
equation~\eqref{eq:Stratonovich stochastic Schrodinger equation}, it is easy to find
\begin{equation}
\alpha_{i}=\sqrt{w_{i}}\sigma_{i}\qquad{\rm and}\qquad \mathcal{H}_{{\rm eff}}=\mathcal{H}_{{\rm QAH}}-(\mathrm{i}/2)\sum_{i}w_{i}.
\end{equation}

In the main text, we have shown that the formal solution of the It\^o stochastic Schr\"odinger equation~\eqref{eq:Ito stochastic Schrodinger equation} is given by a unitary evolution $U(t)$ [see Eq.~\eqref{eq:formal solution}]. To prove that $U(t)$ is indeed the solution of the It\^o
equation, we shall note that
\begin{align}
	U(t+\mathrm{d}t,t)= & \mathrm{e}^{-\mathrm{i}[\mathcal{H}_{{\rm QAH}}\mathrm{d}t+\sum_{i}\sqrt{w_{i}}\sigma_{i}\mathrm{d}W_{i}(t)]}\nonumber\\
	= & \sum_{n=0}^{\infty}\frac{(-\mathrm{i})^{n}}{n!}[\mathcal{H}_{{\rm QAH}}\mathrm{d}t+\sum_{i}\sqrt{w_{i}}\sigma_{i}\mathrm{d}W_{i}(t)]^{n}\nonumber\\
	= & \mathbf{1}-\mathrm{i}[\mathcal{H}_{{\rm eff}}\mathrm{d}t+\sum_{i}\sqrt{w_{i}}\sigma_{i}\mathrm{d}W_{i}(t)],
\end{align}
where the terms other than $\mathrm{d}t$ and $\mathrm{d}W_{i}\mathrm{d}W_{i}=\mathrm{d}t$ vanish according to the It\^o rules. Thus we recover the It\^o stochastic Schr\"odinger equation, i.e. $\mathrm{d}U(t)=U(t+\mathrm{d}t)-U(t)=-\mathrm{i}[\mathcal{H}_{{\rm eff}}\mathrm{d}t+\sum_{i}\sqrt{w_{i}}\sigma_{i}\mathrm{d}W_{i}(t)]U(t)$.

\noindent\textbf{Non-Hermitian dissipative quantum dynamics}.
We now consider the equation of motion for the stochastic density
operator $\varrho(t)=\vert\psi(t)\rangle\langle\psi(t)\vert$, namely
\begin{align}
\mathrm{d}\varrho(t)= & U(t+\mathrm{d}t,t)\varrho(t)U^{\dagger}(t+\mathrm{d}t,t)-\varrho(t)\nonumber\\
 =& -\mathrm{i}\mathcal{H}_{{\rm eff}}\varrho(t)\mathrm{d}t+\mathrm{i}\varrho(t)\mathcal{H}_{{\rm eff}}^{\dagger}\mathrm{d}t+\sum_{i}w_{i}\sigma_{i}\varrho(t)\sigma_{i}\mathrm{d}t\nonumber\\
 & +\mathrm{i}\sum_{i}[\varrho(t),\sqrt{w}_{i}\sigma_{i}]\mathrm{d}W_{i}(t).
\end{align}
Since the increments $\mathrm{d}W_{i}(t)$ are independent of $\varrho(t)$ in the It\^o form, after average over different noise configurations the last term vanishes and we arrive at the Lindblad master equation \eqref{eq:non-Hermitian dissipative quantum dynamics} for the stochastic averaged density matrix
$\rho(t)\equiv\llangle\varrho(t)\rrangle_{{\rm noise}}$,
which describes the non-Hermitian dissipative quantum dynamics.

\noindent\textbf{Stochastically averaged spin dynamics}.\label{sec:Stochastically averaged spin dynamics}
In this section, we show the stochastically averaged spin dynamics. According to the master equation \eqref{eq:non-Hermitian dissipative quantum dynamics}, the stochastically averaged spin polarization $\mathbf{s}(\mathbf{k},t)$ is governed by the equation of motion
\begin{equation}
\partial_{t}\mathbf{s}(\mathbf{k},t)=\mathcal{L}(\mathbf{k})\mathbf{s}(\mathbf{k},t)
\end{equation}
with the Liouvillian superoperator
\begin{equation}
\mathcal{L}(\mathbf{k})=2\begin{bmatrix}-w_{y}-w_{z} & -h_{z}(\mathbf{k}) & h_{y}(\mathbf{k})\\
h_{z}(\mathbf{k}) & -w_{x}-w_{z} & -h_{x}(\mathbf{k})\\
-h_{y}(\mathbf{k}) & h_{x}(\mathbf{k}) & -w_{x}-w_{y}
\end{bmatrix}.
\end{equation}
The solution to this dissipative quantum dynamics can be written as
\begin{align}
\mathbf{s}(\mathbf{k},t)= & \mathbf{s}_{0}(\mathbf{k})\mathrm{e}^{-\lambda_{0}(\mathbf{k})t}+\mathbf{s}_{+}(\mathbf{k})\mathrm{e}^{-[\lambda_{1}(\mathbf{k})+\mathrm{i}\omega(\mathbf{k})]t}\nonumber\\
& +\mathbf{s}_{-}(\mathbf{k})\mathrm{e}^{-[\lambda_{1}(\mathbf{k})-\mathrm{i}\omega(\mathbf{k})]t},
\end{align}
with the coefficients $\mathbf{s}_{\alpha}(\mathbf{k})=[\mathbf{s}_{\alpha}^{L}(\mathbf{k})\cdot\mathbf{s}(\mathbf{k},0)]\mathbf{s}_{\alpha}^{R}$ for $\alpha=0,\pm$. Here $\mathbf{s}_{\alpha}^{L(R)}$ satisfying $\mathbf{s}_{\alpha}^{L}(\mathbf{k})\cdot\mathbf{s}_{\beta}^{R}(\mathbf{k})=\delta_{\alpha\beta}$ are the left (right) eigenvectors of the Liouvillian superoperator $\mathcal{L}^{T}\mathbf{s}_{\alpha}^{L}=-\lambda_{\alpha}\mathbf{s}_{\alpha}^{L}$, $\mathcal{L}\mathbf{s}_{\alpha}^{R}=-\lambda_{\alpha}\mathbf{s}_{\alpha}^{R}$ with eigenvalues $\lambda_{0}$ and $\lambda_{\pm}=\lambda_{1}\pm\mathrm{i}\omega$, respectively. The oscillation frequency is denoted as $\omega$.

In experiments, the coefficients $\mathbf{s}_{\alpha}$, decay rates $\lambda_{0,1}$, and oscillation frequency can be extracted by fitting the experimental data. By ignoring $\lambda_{0,1}$, we obtain the rescaled spin polarization $\tilde{\mathbf{s}}(\mathbf{k},t)$.


\begin{figure}
	\includegraphics[width=\columnwidth]{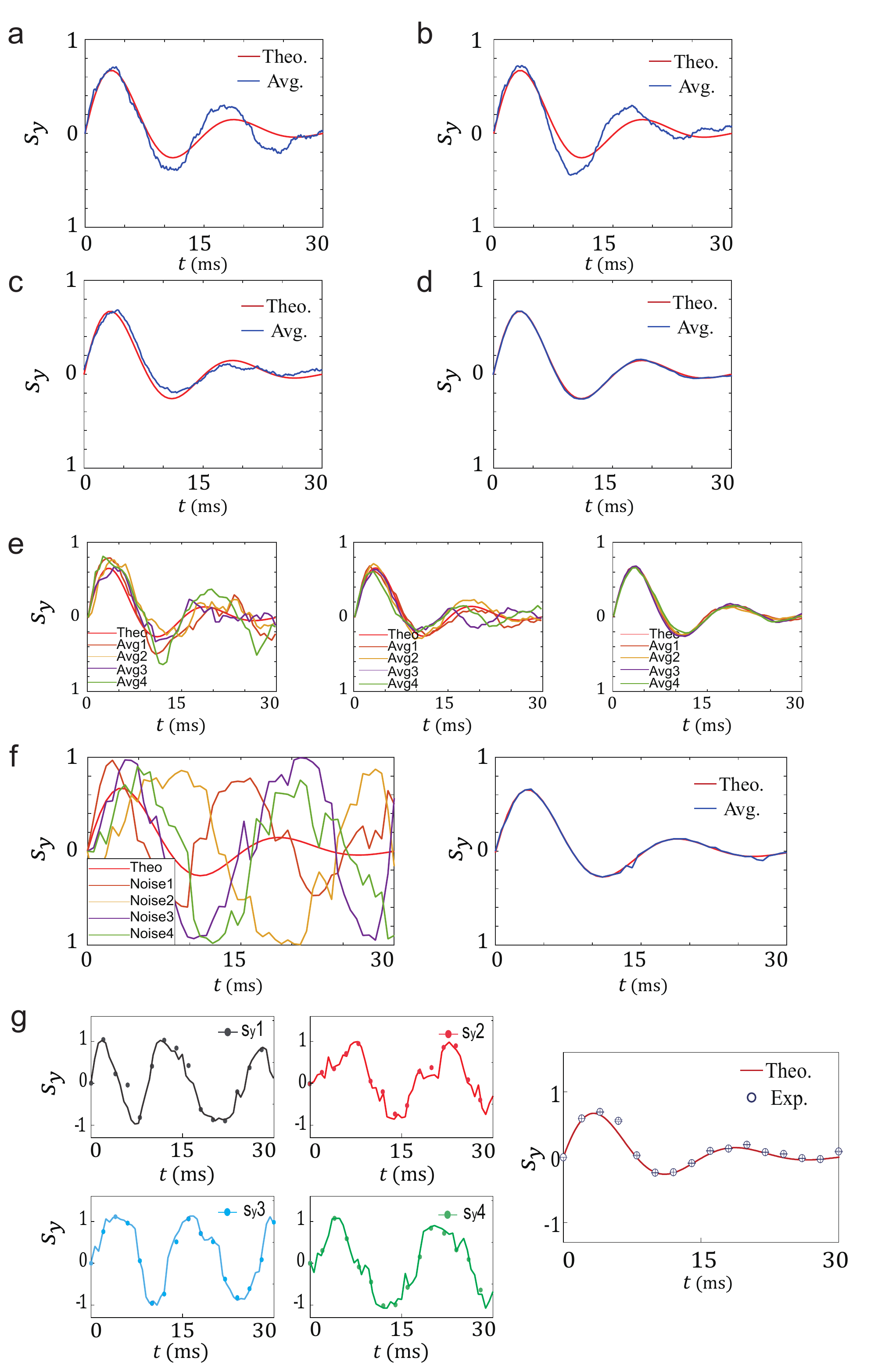}
	\caption{\textbf{Spin polarization produced by solving the master equation and by stochastic average over different number of noise configurations.} $k_x= -1.286, k_y = -0.257$. \textbf{a-d} show results from 50, 100, 200, and 5,000 noise configurations, respectively. \textbf{e} show results from 5, 50, 500 noise configurations, respectively. Each panel contains four average results from the same number of configurations. \textbf{f} Each $\langle \boldsymbol{\upsigma}(\mathbf{k},t) \rangle$ of four pre-simulated noise configurations (left panel). $\llangle\langle \boldsymbol{\upsigma}(\mathbf{k},t) \rangle\rrangle_{\mathrm{noise}}$ averaged by these four pre-simulated noise configurations (right panel). \textbf{g} Experimental and theoretical results of four configurations of noise obtained by pre-simulation. The left four small panels show the experimental and theoretical results of each noise configuration. The line represents the simulation value, and the dot represents the experimental results. The right panel shows the average results from these four noise configurations. \label{Figure8}}	
\end{figure}

\begin{figure}
		\centering
		\includegraphics[width=\columnwidth]{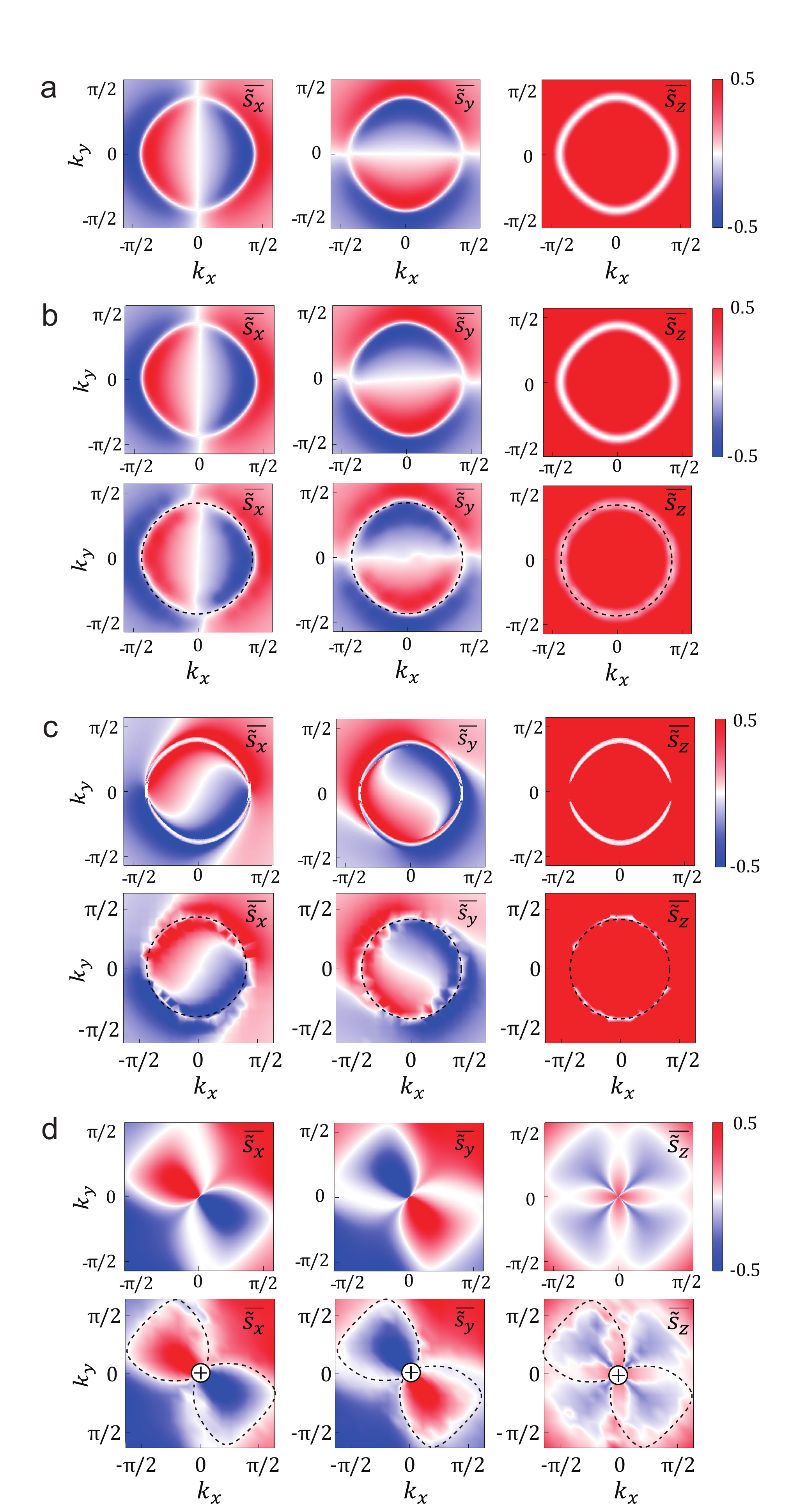}
		\caption{\textbf{Numerical results vs. experimental results.} \textbf{a} 2D time-averaged spin texture of theoretical numerical simulation results  in momentum space $k_x, k_y \in [-1.8, 1.8]$ in the absence of noise, where the ideal BIS momenta can be outlined according to the zero values. \textbf{b} Time-averaged spin texture of theoretical numerical simulation results and experimental results in the presence of noise with noise strength $w_x=0.05\xi_0$, $w_y=0$ $w_z=0.01\xi_0$. The three pictures above are numerical results. The three pictures below are the experimental results obtained by discretizing the momentum space $k_x, k_y \in [-1.8, 1.8]$ into a $15\times 15$ lattice. \textbf{c} Time-averaged spin texture in the presence of noise with noise strength $w_x=0.1\xi_0$, $w_y=0.05\xi_0$, and $w_z=0.45\xi_0$. The pictures above are the numerical results. The pictures below are the experimental results obtained by discretizing the momentum space $k_x, k_y \in [-2, 2]$. \textbf{d} Time-averaged spin texture in the presence of noise with noise strength increase to $w_x=1.6\xi_0$, $w_y=0$, and $w_z=0.8\xi_0$. The pictures above are the numerical results. The pictures below are the experimental results obtained by discretizing the momentum space $k_x, k_y \in [-2, 2]$.\label{Figure9}}	
\end{figure}

\begin{figure}
	\includegraphics[scale=0.6]{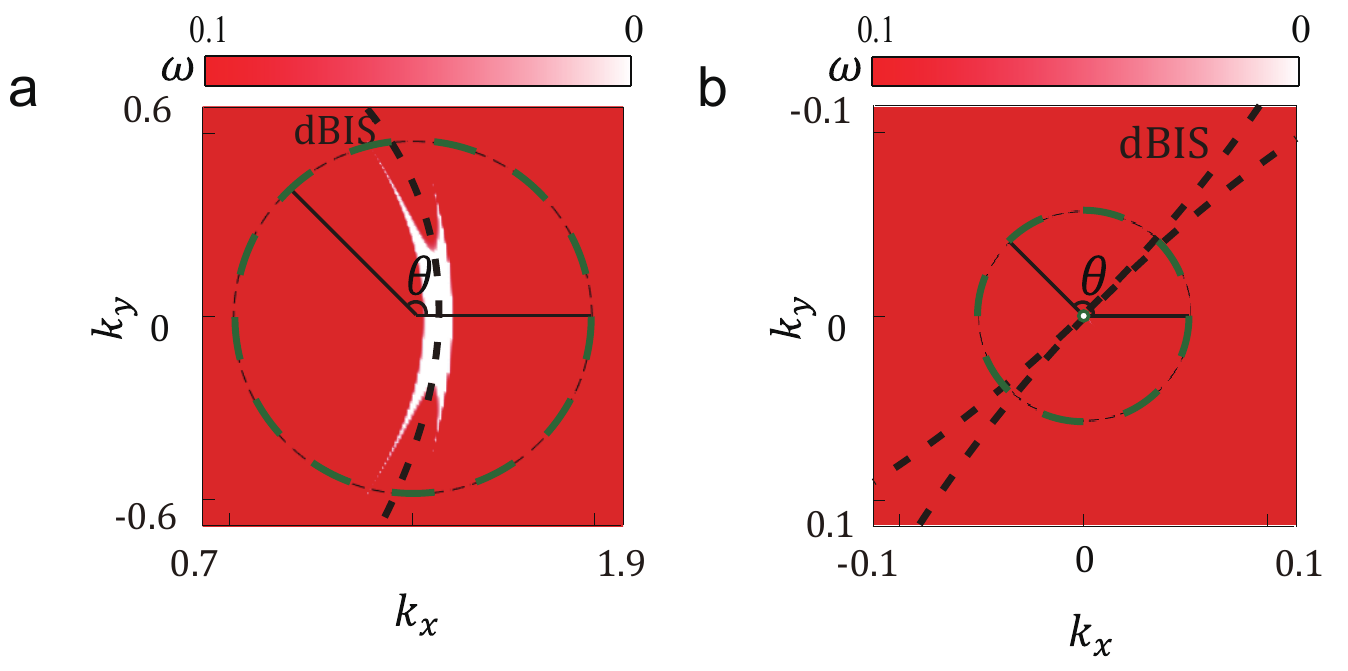}
	\caption{\textbf{Numerical result of exceptional points.} \textbf{a} Exceptional points touch the dBIS (black line) for type-I transition. The noise strength is $w_x=0.1t_0$, $w_y=0.05t_0$, $w_z=0.45t_0$. \textbf{b} Exceptional point (white dot) emerges at the charge momentum (green dot) at $\mathbf{k}=0$, to which the dBIS connects for type-II transition. The noise strength is $w_x=1.6t_0$, $w_y=0t_0$, $w_z=0.8t_0$. \label{Figure10}}	
\end{figure}

%

\noindent\textbf{NMR sample}.
The experiment is performed on the nuclear magnetic
resonance processor (NMR). The sample we used is the $^{13}\text{C-labeled}$ chloroform dissolved in $\text{acetone}-d6$. 
 The $^{13}\text{C}$ spin is used as the working qubit, which is controlled by radio-frequency (RF) fields. The $^{1}\text{H}$ is decoupled throughout the experiment by Overhauser effect which can enhance the signal strength of $^{13}\text{C}$. 

\noindent\textbf{Overhauser effect}.
\label{sec:Overhauser effect}
Applying a weak RF field at the Larmor frequency of one nuclear spin for a sufficient duration may enhance
the longitudinal magnetization of the others, this is the steady-state nuclear Overhauser effect (NOE). In modern NMR, the steady-state NOE is mainly exploited in heteronuclear spin systems, where the enhancement of magnetization is useful and dramatic.

For an ensemble of heteronuclear systems made up with a nuclei $\mathrm{I}$ with gyromagnetic ratio $\gamma_{\mathrm{I}}$ and a nuclei $\mathrm{S}$ with gyromagnetic ratio $\gamma_{\mathrm{S}}$, with $|\gamma_{\mathrm{I}}|>|\gamma_{\mathrm{S}}|$, the thermal equilibrium state of the heteronuclear system is
\begin{equation}
\hat{\rho}^{eq}=\frac{1}{4}\hat{1}+\frac{1}{4}\beta_{\mathrm{I}}\hat{\mathrm{I}}_z+\frac{1}{4}\beta_{\mathrm{S}}\hat{\mathrm{S}}_z,
\label{thermal}
\end{equation}
where $\beta_{\mathrm{I}}/\beta_{\mathrm{S}}=\gamma_{\mathrm{I}}/\gamma_{\mathrm{S}}$, $\frac{1}{4}\hat{\mathrm{I}}_z=\hat{\sigma_z}\otimes \hat{1}$, $\frac{1}{4}\hat{\mathrm{S}}_z=\hat{1}\otimes\hat{\sigma_z}$. Assume that a continuous RF field is applied at the $\mathrm{I}$-spin Larmor frequency, inducing transitions across two pairs of energy levels. After sufficient time, the RF field equalizes the populations across the irradiated transitions. At that time, the populations settle into steady-state values, which do not change any more, as long as the RF field is left on. The steady-state spin density operator is
\begin{equation}
\hat{\rho}^{ss}=\frac{1}{4}\hat{1}+\epsilon_{\text{NOE}}\frac{1}{4}\beta_{\mathrm{S}}\hat{\mathrm{S}}_z.
\end{equation}
By comparing with thermal equilibrium Eq.~(\ref{thermal}), the $\mathrm{S}$-spin magnetization is enhanced by factor $\epsilon_{\text{NOE}}$. For our experiment $\mathrm{I}=^{1}\text{H}$ and $\mathrm{S}=^{13}\text{C}$.

\noindent\textbf{Noise configurations}.
For the stochastic average, it is clear that the more noise configurations are considered, the more reasonable result we obtain, as shown in Fig.~\ref{Figure8}. On the other hand, the large number of noise configurations takes a lot of time. We have performed numerical simulations, and find that the average of 5,000 noise configurations can precisely approximate the non-Hermitian dissipative quantum dynamics; see Fig.~\ref{Figure8}d. However, in NMR experiments, as the relaxation time is in the magnitude of seconds, a complete implementation of all 5,000 noise configurations requires an extremely long running time that we cannot afford.

An alternative method to solve the issue is to reduce the number of noise configurations by numerical simulation prior to the implementation of experiments. We test different number of noise configurations, and plot their average dynamics in comparison with the ideal dynamics of the non-Hermitian Hamiltonian; see Fig.~\ref{Figure8}a-d. The simulated results show that with the increase of the number of noise configurations, the stochastic averaged spin polarization $\llangle\langle \boldsymbol{\upsigma}(\mathbf{k},t) \rangle\rrangle_{\mathrm{noise}}$ would eventually approach to the spin polarization $\mathbf{s}(\mathbf{k},t)$ solved by the Lindblad master equation \cite{Zhang2021noise}. The opposite is that with the decrease of the number of noise configurations, the performance of the approximation becomes more fluctuating [Fig.~\ref{Figure8}e]. But the $\llangle\langle \boldsymbol{\upsigma}(\mathbf{k},t) \rangle\rrangle_{\mathrm{noise}}$ always fluctuates above and below the theoretical spin polarization $\mathbf{s}(\mathbf{k},t)$. After a sufficient number of averaging, the stochastic averaged spin polarization that in the opposite side of theoretical value will be offset by each other. We randomly generated 5,000 noise configurations $N(t_{n})$ that satisfy the normal distribution and separate these noise configurations into two subgroups in which the noise has opposite effect on $\llangle\langle \boldsymbol{\upsigma}(\mathbf{k},t) \rangle\rrangle_{\mathrm{noise}}$. Then we use numerical simulations to select two noise configurations from these two subgroups respectively such that the $\llangle\langle \boldsymbol{\upsigma}(\mathbf{k},t) \rangle\rrangle_{\mathrm{noise}}$ obtained from these four noise configurations precisely approximate the one obtained from the 5,000 configurations [Fig.~\ref{Figure8}f]. From experimental results and the corresponding fidelities, it can be concluded that the experiment is in excellent accordance with the simulations. And the theory and experiment results of each group of noise are in good agreement [Fig.~\ref{Figure8}g] So, it is somehow reasonable to utilize four noise configurations to replace a full description of the non-Hermitian dynamics under 5,000 noise configurations.
We would like to emphasize that the above numerical simulations to reduce the number of noise configurations does not affect the applicability of the method. In many other quantum systems such as the superconducting circuits or nitrogen-vacancy centres in diamond, the implementation of experiments takes much shorter time, so they can realize the stochastic average with a larger number of noise configurations.

\noindent\textbf{Experimental results vs. theoretical results}.\label{sec:Experimental vs theoretical}
In this section, we show the agreement of our experimental results with the theoretical calculations. In Fig.~\ref{Figure9}, we compare the experimental spin textures with theoretical ones. Although the resolution of experimental data is lower than that of numerical calculations, the experimental results and the theoretical simulations reach the same conclusion. In Fig.~\ref{Figure10}, we show the numerical calculations for exceptional points and the corresponding winding numbers, which are consistent with our experimental results (see Fig.~\ref{Figure7}).

\section*{DATA AVAILABILITY}
The datasets generated during and/or analysed during the current study are available from the corresponding author on reasonable request.

\section*{ACKNOWLEDGEMENTS}
This work is supported by the National Key Research and Development Program of China (2019YFA0308100, 2021YFA1400900), the National Natural Science Foundation of China (12075110, 11825401, 11975117, 11905099, 11875159, and U1801661), the Guangdong Basic and Applied Basic Research Foundation (2019A1515011383, 2021B1515020070), the Guangdong International Collaboration Program (2020A0505100001), the Science, Technology and Innovation Commission of Shenzhen Municipality (ZDSYS20170303165926217, KQTD20190929173815000, JCYJ20200109140803865, JCYJ20170412152620376, RCYX20200714114522109, and JCYJ20180302174036418), the Open Project of Shenzhen Institute of Quantum Science and Engineering (Grant No.SIQSE202003), the Pengcheng Scholars, the Guangdong Innovative and Entrepreneurial Research Team Program (2019ZT08C044), and the Guangdong Provincial Key Laboratory (2019B121203002). L.Z. also acknowledges support from Agencia Estatal de Investigaci{\'o}n (the R\&D project CEX2019-000910-S, funded by MCIN/AEI/10.13039/501100011033, Plan National FIDEUA PID2019-106901GB-I00, FPI), Fundaci{\'o} Privada Cellex, Fundaci{\'o} Mir-Puig, and Generalitat de Catalunya (AGAUR Grant No. 2017 SGR 1341, CERCA program).

\section*{COMPETING INTERESTS}
The authors declare no competing interests.
 
\section*{AUTHOR CONTRIBUTIONS}
D.L. and X.L. supervised the experiments. L.Z. and X.L. elaborated the theoretical framework. Z.L. and X.L. wrote the computer code and accomplished the NMR experiments. All authors analyzed the data, discussed the results, and wrote the manuscript. Z.L., L.Z. and X.L. who made equal contributions to this work are considered "co-first authors".
 
\section*{CORRESPONDENCE}
Correspondence should be addressed to D.L.~(email: ludw@sustech.edu.cn) or X.L.~(email: xiongjunliu@pku.edu.cn) or T.X.~(email: xint@sustech.edu.cn).

\providecommand{\noopsort}[1]{}\providecommand{\singleletter}[1]{#1}%


\begin{thebibliography}{10}
\expandafter\ifx\csname url\endcsname\relax
  \def\url#1{\texttt{#1}}\fi
\expandafter\ifx\csname urlprefix\endcsname\relax\def\urlprefix{URL }\fi
\providecommand{\bibinfo}[2]{#2}
\providecommand{\eprint}[2][]{\url{#2}}

\bibitem{landau1999statistical}
\bibinfo{author}{Landau, L.} \& \bibinfo{author}{Lifshitz, E.}
\newblock \bibinfo{title}{Statistical physics, course theoretical phys., vol.
  5} (\bibinfo{year}{1999}).

\bibitem{hasan2010colloquium}
\bibinfo{author}{Hasan, M.~Z.} \& \bibinfo{author}{Kane, C.~L.}
\newblock \bibinfo{title}{Colloquium: topological insulators}.
\newblock \emph{\bibinfo{journal}{Rev. Mod. Phys.}}
  \textbf{\bibinfo{volume}{82}}, \bibinfo{pages}{3045} (\bibinfo{year}{2010}).

\bibitem{qi2011topological}
\bibinfo{author}{Qi, X.-L.} \& \bibinfo{author}{Zhang, S.-C.}
\newblock \bibinfo{title}{Topological insulators and superconductors}.
\newblock \emph{\bibinfo{journal}{Rev. Mod. Phys.}}
  \textbf{\bibinfo{volume}{83}}, \bibinfo{pages}{1057} (\bibinfo{year}{2011}).

\bibitem{chiu2016classification}
\bibinfo{author}{Chiu, C.-K.}, \bibinfo{author}{Teo, J.~C.},
  \bibinfo{author}{Schnyder, A.~P.} \& \bibinfo{author}{Ryu, S.}
\newblock \bibinfo{title}{Classification of topological quantum matter with
  symmetries}.
\newblock \emph{\bibinfo{journal}{Rev. Mod. Phys.}}
  \textbf{\bibinfo{volume}{88}}, \bibinfo{pages}{035005}
  (\bibinfo{year}{2016}).

\bibitem{ando2015topological}
\bibinfo{author}{Ando, Y.} \& \bibinfo{author}{Fu, L.}
\newblock \bibinfo{title}{Topological crystalline insulators and topological
  superconductors: From concepts to materials}.
\newblock \emph{\bibinfo{journal}{Annu. Rev. Condens. Matter Phys.}}
  \textbf{\bibinfo{volume}{6}}, \bibinfo{pages}{361--381}
  (\bibinfo{year}{2015}).

\bibitem{sato2017topological}
\bibinfo{author}{Sato, M.} \& \bibinfo{author}{Ando, Y.}
\newblock \bibinfo{title}{Topological superconductors: a review}.
\newblock \emph{\bibinfo{journal}{Rep. Prog. Phys.}}
  \textbf{\bibinfo{volume}{80}}, \bibinfo{pages}{076501}
  (\bibinfo{year}{2017}).

\bibitem{xie2021higher}
\bibinfo{author}{Xie, B.} \emph{et~al.}
\newblock \bibinfo{title}{Higher-order band topology}.
\newblock \emph{\bibinfo{journal}{Nat. Rev. Phys.}}
  \textbf{\bibinfo{volume}{3}}, \bibinfo{pages}{520--532}
  (\bibinfo{year}{2021}).

\bibitem{Konig2007}
\bibinfo{author}{Konig, M.} \emph{et~al.}
\newblock \bibinfo{title}{Quantum spin hall insulator state in hgte quantum
  wells}.
\newblock \emph{\bibinfo{journal}{Science}} \textbf{\bibinfo{volume}{318}},
  \bibinfo{pages}{766--770} (\bibinfo{year}{2007}).

\bibitem{Chang2013}
\bibinfo{author}{Chang, C.-Z.} \emph{et~al.}
\newblock \bibinfo{title}{Experimental observation of the quantum anomalous
  hall effect in a magnetic topological insulator}.
\newblock \emph{\bibinfo{journal}{Science}} \textbf{\bibinfo{volume}{340}},
  \bibinfo{pages}{167--170} (\bibinfo{year}{2013}).

\bibitem{He2017}
\bibinfo{author}{He, Q.~L.} \emph{et~al.}
\newblock \bibinfo{title}{Chiral majorana fermion modes in a quantum anomalous
  hall insulator--superconductor structure}.
\newblock \emph{\bibinfo{journal}{Science}} \textbf{\bibinfo{volume}{357}},
  \bibinfo{pages}{294--299} (\bibinfo{year}{2017}).

\bibitem{Hsieh2008}
\bibinfo{author}{Hsieh, D.} \emph{et~al.}
\newblock \bibinfo{title}{A topological dirac insulator in a quantum spin hall
  phase}.
\newblock \emph{\bibinfo{journal}{Nature}} \textbf{\bibinfo{volume}{452}},
  \bibinfo{pages}{970--974} (\bibinfo{year}{2008}).

\bibitem{Chen2009}
\bibinfo{author}{Chen, Y.} \emph{et~al.}
\newblock \bibinfo{title}{Experimental realization of a three-dimensional
  topological insulator, bi2te3}.
\newblock \emph{\bibinfo{journal}{Science}} \textbf{\bibinfo{volume}{325}},
  \bibinfo{pages}{178--181} (\bibinfo{year}{2009}).

\bibitem{Xia2009}
\bibinfo{author}{Xia, Y.} \emph{et~al.}
\newblock \bibinfo{title}{Observation of a large-gap topological-insulator
  class with a single dirac cone on the surface}.
\newblock \emph{\bibinfo{journal}{Nat. Phys.}} \textbf{\bibinfo{volume}{5}},
  \bibinfo{pages}{398--402} (\bibinfo{year}{2009}).

\bibitem{Vajna2015}
\bibinfo{author}{Vajna, S.} \& \bibinfo{author}{D{\'o}ra, B.}
\newblock \bibinfo{title}{Topological classification of dynamical phase
  transitions}.
\newblock \emph{\bibinfo{journal}{Phys. Rev. B}} \textbf{\bibinfo{volume}{91}},
  \bibinfo{pages}{155127} (\bibinfo{year}{2015}).

\bibitem{Caio2015}
\bibinfo{author}{Caio, M.}, \bibinfo{author}{Cooper, N.~R.} \&
  \bibinfo{author}{Bhaseen, M.}
\newblock \bibinfo{title}{Quantum quenches in chern insulators}.
\newblock \emph{\bibinfo{journal}{Phys. Rev. Lett.}}
  \textbf{\bibinfo{volume}{115}}, \bibinfo{pages}{236403}
  (\bibinfo{year}{2015}).

\bibitem{Budich2016}
\bibinfo{author}{Budich, J.~C.} \& \bibinfo{author}{Heyl, M.}
\newblock \bibinfo{title}{Dynamical topological order parameters far from
  equilibrium}.
\newblock \emph{\bibinfo{journal}{Phys. Rev. B}} \textbf{\bibinfo{volume}{93}},
  \bibinfo{pages}{085416} (\bibinfo{year}{2016}).

\bibitem{Wilson2016}
\bibinfo{author}{Wilson, J.~H.}, \bibinfo{author}{Song, J.~C.} \&
  \bibinfo{author}{Refael, G.}
\newblock \bibinfo{title}{Remnant geometric hall response in a quantum quench}.
\newblock \emph{\bibinfo{journal}{Phys. Rev. Lett.}}
  \textbf{\bibinfo{volume}{117}}, \bibinfo{pages}{235302}
  (\bibinfo{year}{2016}).

\bibitem{Gong2018}
\bibinfo{author}{Gong, Z.} \& \bibinfo{author}{Ueda, M.}
\newblock \bibinfo{title}{Topological entanglement-spectrum crossing in quench
  dynamics}.
\newblock \emph{\bibinfo{journal}{Phys. Rev. Lett.}}
  \textbf{\bibinfo{volume}{121}}, \bibinfo{pages}{250601}
  (\bibinfo{year}{2018}).

\bibitem{Flaschner2018}
\bibinfo{author}{Fl{\"a}schner, N.} \emph{et~al.}
\newblock \bibinfo{title}{Observation of dynamical vortices after quenches in a
  system with topology}.
\newblock \emph{\bibinfo{journal}{Nat. Phys.}} \textbf{\bibinfo{volume}{14}},
  \bibinfo{pages}{265--268} (\bibinfo{year}{2018}).

\bibitem{Song2018}
\bibinfo{author}{Song, B.} \emph{et~al.}
\newblock \bibinfo{title}{Observation of symmetry-protected topological band
  with ultracold fermions}.
\newblock \emph{\bibinfo{journal}{Sci. Adv.}} \textbf{\bibinfo{volume}{4}},
  \bibinfo{pages}{eaao4748} (\bibinfo{year}{2018}).

\bibitem{Qiu2019}
\bibinfo{author}{Qiu, X.}, \bibinfo{author}{Deng, T.-S.}, \bibinfo{author}{Hu,
  Y.}, \bibinfo{author}{Xue, P.} \& \bibinfo{author}{Yi, W.}
\newblock \bibinfo{title}{Fixed points and dynamic topological phenomena in a
  parity-time-symmetric quantum quench}.
\newblock \emph{\bibinfo{journal}{iScience}} \textbf{\bibinfo{volume}{20}},
  \bibinfo{pages}{392--401} (\bibinfo{year}{2019}).

\bibitem{Huang2020}
\bibinfo{author}{Huang, B.} \& \bibinfo{author}{Liu, W.~V.}
\newblock \bibinfo{title}{Floquet higher-order topological insulators with
  anomalous dynamical polarization}.
\newblock \emph{\bibinfo{journal}{Phys. Rev. Lett.}}
  \textbf{\bibinfo{volume}{124}}, \bibinfo{pages}{216601}
  (\bibinfo{year}{2020}).

\bibitem{Unal2020}
\bibinfo{author}{{\"U}nal, F.~N.}, \bibinfo{author}{Bouhon, A.} \&
  \bibinfo{author}{Slager, R.-J.}
\newblock \bibinfo{title}{Topological euler class as a dynamical observable in
  optical lattices}.
\newblock \emph{\bibinfo{journal}{Phys. Rev. Lett.}}
  \textbf{\bibinfo{volume}{125}}, \bibinfo{pages}{053601}
  (\bibinfo{year}{2020}).

\bibitem{Hu2020}
\bibinfo{author}{Hu, H.} \& \bibinfo{author}{Zhao, E.}
\newblock \bibinfo{title}{Topological invariants for quantum quench dynamics
  from unitary evolution}.
\newblock \emph{\bibinfo{journal}{Phys. Rev. Lett.}}
  \textbf{\bibinfo{volume}{124}}, \bibinfo{pages}{160402}
  (\bibinfo{year}{2020}).

\bibitem{Hu2020Hopf}
\bibinfo{author}{Hu, H.}, \bibinfo{author}{Yang, C.} \& \bibinfo{author}{Zhao,
  E.}
\newblock \bibinfo{title}{Quench dynamics of hopf insulators}.
\newblock \emph{\bibinfo{journal}{Phys. Rev. B}}
  \textbf{\bibinfo{volume}{101}}, \bibinfo{pages}{155131}
  (\bibinfo{year}{2020}).

\bibitem{Lu2020}
\bibinfo{author}{Lu, Y.-H.}, \bibinfo{author}{Wang, B.-Z.} \&
  \bibinfo{author}{Liu, X.-J.}
\newblock \bibinfo{title}{Ideal weyl semimetal with 3d spin-orbit coupled
  ultracold quantum gas}.
\newblock \emph{\bibinfo{journal}{Sci. Bull.}} \textbf{\bibinfo{volume}{65}},
  \bibinfo{pages}{2080--2085} (\bibinfo{year}{2020}).

\bibitem{Wang2021}
\bibinfo{author}{Wang, Z.-Y.} \emph{et~al.}
\newblock \bibinfo{title}{Realization of an ideal weyl semimetal band in a
  quantum gas with 3d spin-orbit coupling}.
\newblock \emph{\bibinfo{journal}{Science}} \textbf{\bibinfo{volume}{372}},
  \bibinfo{pages}{271--276} (\bibinfo{year}{2021}).

\bibitem{Mizoguchi2021}
\bibinfo{author}{Mizoguchi, T.}, \bibinfo{author}{Kuno, Y.} \&
  \bibinfo{author}{Hatsugai, Y.}
\newblock \bibinfo{title}{Detecting bulk topology of quadrupolar phase from
  quench dynamics}.
\newblock \emph{\bibinfo{journal}{Phys. Rev. Lett.}}
  \textbf{\bibinfo{volume}{126}}, \bibinfo{pages}{016802}
  (\bibinfo{year}{2021}).

\bibitem{Zhang2018}
\bibinfo{author}{Zhang, L.}, \bibinfo{author}{Zhang, L.}, \bibinfo{author}{Niu,
  S.} \& \bibinfo{author}{Liu, X.-J.}
\newblock \bibinfo{title}{Dynamical classification of topological quantum
  phases}.
\newblock \emph{\bibinfo{journal}{Sci. Bull.}} \textbf{\bibinfo{volume}{63}},
  \bibinfo{pages}{1385--1391} (\bibinfo{year}{2018}).

\bibitem{Zhang2019}
\bibinfo{author}{Zhang, L.}, \bibinfo{author}{Zhang, L.} \&
  \bibinfo{author}{Liu, X.-J.}
\newblock \bibinfo{title}{Dynamical detection of topological charges}.
\newblock \emph{\bibinfo{journal}{Phys. Rev. A}} \textbf{\bibinfo{volume}{99}},
  \bibinfo{pages}{053606} (\bibinfo{year}{2019}).

\bibitem{Zhang2020Floquet}
\bibinfo{author}{Zhang, L.}, \bibinfo{author}{Zhang, L.} \&
  \bibinfo{author}{Liu, X.-J.}
\newblock \bibinfo{title}{Unified theory to characterize floquet topological
  phases by quench dynamics}.
\newblock \emph{\bibinfo{journal}{Phys. Rev. Lett.}}
  \textbf{\bibinfo{volume}{125}}, \bibinfo{pages}{183001}
  (\bibinfo{year}{2020}).

\bibitem{Yu2021}
\bibinfo{author}{Yu, X.-L.} \emph{et~al.}
\newblock \bibinfo{title}{Quantum dynamical characterization and simulation of
  topological phases with high-order band inversion surfaces}.
\newblock \emph{\bibinfo{journal}{PRX Quantum}} \textbf{\bibinfo{volume}{2}},
  \bibinfo{pages}{020320} (\bibinfo{year}{2021}).

\bibitem{Li2021}
\bibinfo{author}{Li, L.}, \bibinfo{author}{Zhu, W.} \& \bibinfo{author}{Gong,
  J.}
\newblock \bibinfo{title}{Direct dynamical characterization of higher-order
  topological phases with nested band inversion surfaces}.
\newblock \emph{\bibinfo{journal}{Sci. Bull.}} \textbf{\bibinfo{volume}{66}},
  \bibinfo{pages}{1502--1510} (\bibinfo{year}{2021}).

\bibitem{Zhang2021AZ}
\bibinfo{author}{Zhang, L.}, \bibinfo{author}{Jia, W.} \& \bibinfo{author}{Liu,
  X.-J.}
\newblock \bibinfo{title}{Universal topological quench dynamics for
  $\mathbb{Z}_{2}$ topological phases}.
\newblock \emph{\bibinfo{journal}{Sci. Bull.}}  (\bibinfo{year}{2022}).

\bibitem{Sun2018}
\bibinfo{author}{Sun, W.} \emph{et~al.}
\newblock \bibinfo{title}{Uncover topology by quantum quench dynamics}.
\newblock \emph{\bibinfo{journal}{Phys. Rev. Lett.}}
  \textbf{\bibinfo{volume}{121}}, \bibinfo{pages}{250403}
  (\bibinfo{year}{2018}).

\bibitem{Yi2019}
\bibinfo{author}{Yi, C.-R.} \emph{et~al.}
\newblock \bibinfo{title}{Observing topological charges and dynamical
  bulk-surface correspondence with ultracold atoms}.
\newblock \emph{\bibinfo{journal}{Phys. Rev. Lett.}}
  \textbf{\bibinfo{volume}{123}}, \bibinfo{pages}{190603}
  (\bibinfo{year}{2019}).

\bibitem{Wang2019}
\bibinfo{author}{Wang, Y.} \emph{et~al.}
\newblock \bibinfo{title}{Experimental observation of dynamical bulk-surface
  correspondence in momentum space for topological phases}.
\newblock \emph{\bibinfo{journal}{Phys. Rev. A}}
  \textbf{\bibinfo{volume}{100}}, \bibinfo{pages}{052328}
  (\bibinfo{year}{2019}).

\bibitem{Ji2020}
\bibinfo{author}{Ji, W.} \emph{et~al.}
\newblock \bibinfo{title}{Quantum simulation for three-dimensional chiral
  topological insulator}.
\newblock \emph{\bibinfo{journal}{Phys. Rev. Lett.}}
  \textbf{\bibinfo{volume}{125}}, \bibinfo{pages}{020504}
  (\bibinfo{year}{2020}).

\bibitem{Chen2021}
\bibinfo{author}{Chen, B.} \emph{et~al.}
\newblock \bibinfo{title}{Digital quantum simulation of floquet topological
  phases with a solid-state quantum simulator}.
\newblock \emph{\bibinfo{journal}{Photon. Res.}} \textbf{\bibinfo{volume}{9}},
  \bibinfo{pages}{81--87} (\bibinfo{year}{2021}).

\bibitem{Xin2020}
\bibinfo{author}{Xin, T.} \emph{et~al.}
\newblock \bibinfo{title}{Quantum phases of three-dimensional chiral
  topological insulators on a spin quantum simulator}.
\newblock \emph{\bibinfo{journal}{Phys. Rev. Lett.}}
  \textbf{\bibinfo{volume}{125}}, \bibinfo{pages}{090502}
  (\bibinfo{year}{2020}).

\bibitem{Niu2021}
\bibinfo{author}{Niu, J.} \emph{et~al.}
\newblock \bibinfo{title}{Simulation of higher-order topological phases and
  related topological phase transitions in a superconducting qubit}.
\newblock \emph{\bibinfo{journal}{Sci. Bull.}} \textbf{\bibinfo{volume}{66}},
  \bibinfo{pages}{1168--1175} (\bibinfo{year}{2021}).

\bibitem{Breuer2007}
\bibinfo{author}{Breuer, H.-P.}, \bibinfo{author}{Petruccione, F.}
  \emph{et~al.}
\newblock \emph{\bibinfo{title}{The theory of open quantum systems}}
  (\bibinfo{publisher}{Oxford University Press on Demand},
  \bibinfo{year}{2002}).

\bibitem{Ashida2020}
\bibinfo{author}{Ashida, Y.}, \bibinfo{author}{Gong, Z.} \&
  \bibinfo{author}{Ueda, M.}
\newblock \bibinfo{title}{Non-hermitian physics}.
\newblock \emph{\bibinfo{journal}{Adv. Phys.}} \textbf{\bibinfo{volume}{69}},
  \bibinfo{pages}{249--435} (\bibinfo{year}{2020}).

\bibitem{Bergholtz2021}
\bibinfo{author}{Bergholtz, E.~J.}, \bibinfo{author}{Budich, J.~C.} \&
  \bibinfo{author}{Kunst, F.~K.}
\newblock \bibinfo{title}{Exceptional topology of non-hermitian systems}.
\newblock \emph{\bibinfo{journal}{Rev. Mod. Phys.}}
  \textbf{\bibinfo{volume}{93}}, \bibinfo{pages}{015005}
  (\bibinfo{year}{2021}).

\bibitem{Heiss2012}
\bibinfo{author}{Heiss, W.}
\newblock \bibinfo{title}{The physics of exceptional points}.
\newblock \emph{\bibinfo{journal}{J. Phys. A}} \textbf{\bibinfo{volume}{45}},
  \bibinfo{pages}{444016} (\bibinfo{year}{2012}).

\bibitem{Lee2016}
\bibinfo{author}{Lee, T.~E.}
\newblock \bibinfo{title}{Anomalous edge state in a non-hermitian lattice}.
\newblock \emph{\bibinfo{journal}{Phys. Rev. Lett.}}
  \textbf{\bibinfo{volume}{116}}, \bibinfo{pages}{133903}
  (\bibinfo{year}{2016}).

\bibitem{Leykam2017}
\bibinfo{author}{Leykam, D.}, \bibinfo{author}{Bliokh, K.~Y.},
  \bibinfo{author}{Huang, C.}, \bibinfo{author}{Chong, Y.~D.} \&
  \bibinfo{author}{Nori, F.}
\newblock \bibinfo{title}{Edge modes, degeneracies, and topological numbers in
  non-hermitian systems}.
\newblock \emph{\bibinfo{journal}{Phys. Rev. Lett.}}
  \textbf{\bibinfo{volume}{118}}, \bibinfo{pages}{040401}
  (\bibinfo{year}{2017}).

\bibitem{Xu2017}
\bibinfo{author}{Xu, Y.}, \bibinfo{author}{Wang, S.-T.} \&
  \bibinfo{author}{Duan, L.-M.}
\newblock \bibinfo{title}{Weyl exceptional rings in a three-dimensional
  dissipative cold atomic gas}.
\newblock \emph{\bibinfo{journal}{Phys. Rev. Lett.}}
  \textbf{\bibinfo{volume}{118}}, \bibinfo{pages}{045701}
  (\bibinfo{year}{2017}).

\bibitem{Xiong2018}
\bibinfo{author}{Xiong, Y.}
\newblock \bibinfo{title}{Why does bulk boundary correspondence fail in some
  non-hermitian topological models}.
\newblock \emph{\bibinfo{journal}{J. Phys. Commun.}}
  \textbf{\bibinfo{volume}{2}}, \bibinfo{pages}{035043} (\bibinfo{year}{2018}).

\bibitem{Kunst2018}
\bibinfo{author}{Kunst, F.~K.}, \bibinfo{author}{Edvardsson, E.},
  \bibinfo{author}{Budich, J.~C.} \& \bibinfo{author}{Bergholtz, E.~J.}
\newblock \bibinfo{title}{Biorthogonal bulk-boundary correspondence in
  non-hermitian systems}.
\newblock \emph{\bibinfo{journal}{Phys. Rev. Lett.}}
  \textbf{\bibinfo{volume}{121}}, \bibinfo{pages}{026808}
  (\bibinfo{year}{2018}).

\bibitem{Yao2018a}
\bibinfo{author}{Yao, S.}, \bibinfo{author}{Song, F.} \& \bibinfo{author}{Wang,
  Z.}
\newblock \bibinfo{title}{Non-hermitian chern bands}.
\newblock \emph{\bibinfo{journal}{Phys. Rev. Lett.}}
  \textbf{\bibinfo{volume}{121}}, \bibinfo{pages}{136802}
  (\bibinfo{year}{2018}).

\bibitem{Yao2018b}
\bibinfo{author}{Yao, S.} \& \bibinfo{author}{Wang, Z.}
\newblock \bibinfo{title}{Edge states and topological invariants of
  non-hermitian systems}.
\newblock \emph{\bibinfo{journal}{Phys. Rev. Lett.}}
  \textbf{\bibinfo{volume}{121}}, \bibinfo{pages}{086803}
  (\bibinfo{year}{2018}).

\bibitem{Xiao2020}
\bibinfo{author}{Xiao, L.} \emph{et~al.}
\newblock \bibinfo{title}{Non-hermitian bulk--boundary correspondence in
  quantum dynamics}.
\newblock \emph{\bibinfo{journal}{Nat. Phys.}} \textbf{\bibinfo{volume}{16}},
  \bibinfo{pages}{761--766} (\bibinfo{year}{2020}).

\bibitem{Weidemann2020}
\bibinfo{author}{Weidemann, S.} \emph{et~al.}
\newblock \bibinfo{title}{Topological funneling of light}.
\newblock \emph{\bibinfo{journal}{Science}} \textbf{\bibinfo{volume}{368}},
  \bibinfo{pages}{311--314} (\bibinfo{year}{2020}).

\bibitem{Ghatak2020}
\bibinfo{author}{Ghatak, A.}, \bibinfo{author}{Brandenbourger, M.},
  \bibinfo{author}{Van~Wezel, J.} \& \bibinfo{author}{Coulais, C.}
\newblock \bibinfo{title}{Observation of non-hermitian topology and its
  bulk--edge correspondence in an active mechanical metamaterial}.
\newblock \emph{\bibinfo{journal}{Proc. Natl. Acad. Sci.}}
  \textbf{\bibinfo{volume}{117}}, \bibinfo{pages}{29561--29568}
  (\bibinfo{year}{2020}).

\bibitem{Helbig2020}
\bibinfo{author}{Helbig, T.} \emph{et~al.}
\newblock \bibinfo{title}{Generalized bulk--boundary correspondence in
  non-hermitian topolectrical circuits}.
\newblock \emph{\bibinfo{journal}{Nat. Phys.}} \textbf{\bibinfo{volume}{16}},
  \bibinfo{pages}{747--750} (\bibinfo{year}{2020}).

\bibitem{Wu2019}
\bibinfo{author}{Wu, Y.} \emph{et~al.}
\newblock \bibinfo{title}{Observation of parity-time symmetry breaking in a
  single-spin system}.
\newblock \emph{\bibinfo{journal}{Science}} \textbf{\bibinfo{volume}{364}},
  \bibinfo{pages}{878--880} (\bibinfo{year}{2019}).

\bibitem{Zhang2020}
\bibinfo{author}{Zhang, W.} \emph{et~al.}
\newblock \bibinfo{title}{Observation of non-hermitian topology with nonunitary
  dynamics of solid-state spins}.
\newblock \emph{\bibinfo{journal}{Phys. Rev. Lett.}}
  \textbf{\bibinfo{volume}{127}}, \bibinfo{pages}{090501}
  (\bibinfo{year}{2021}).

\bibitem{Gardiner2004}
\bibinfo{author}{Gardiner, C.} \& \bibinfo{author}{Zoller, P.}
\newblock \bibinfo{title}{Quantum noise, springer-verlag, berlin (2000)}
  (\bibinfo{year}{2004}).

\bibitem{Gardiner2014}
\bibinfo{author}{Gardiner, C.} \& \bibinfo{author}{Zoller, P.}
\newblock \emph{\bibinfo{title}{The Quantum World of Ultra-Cold Atoms and Light
  Book I: Foundations of Quantum Optics}}, vol.~\bibinfo{volume}{2}
  (\bibinfo{publisher}{World Scientific Publishing Company},
  \bibinfo{year}{2014}).

\bibitem{Zhang2021noise}
\bibinfo{author}{Zhang, L.}, \bibinfo{author}{Zhang, L.} \&
  \bibinfo{author}{Liu, X.-J.}
\newblock \bibinfo{title}{Quench-induced dynamical topology under dynamical
  noise}.
\newblock \emph{\bibinfo{journal}{Phys. Rev. Research}}
  \textbf{\bibinfo{volume}{3}}, \bibinfo{pages}{013229} (\bibinfo{year}{2021}).

\bibitem{Liu2014}
\bibinfo{author}{Liu, X.-J.}, \bibinfo{author}{Law, K.~T.} \&
  \bibinfo{author}{Ng, T.~K.}
\newblock \bibinfo{title}{Realization of 2d spin-orbit interaction and exotic
  topological orders in cold atoms}.
\newblock \emph{\bibinfo{journal}{Phys. Rev. Lett.}}
  \textbf{\bibinfo{volume}{112}}, \bibinfo{pages}{086401}
  (\bibinfo{year}{2014}).

\bibitem{Wu2016}
\bibinfo{author}{Wu, Z.} \emph{et~al.}
\newblock \bibinfo{title}{Realization of two-dimensional spin-orbit coupling
  for bose-einstein condensates}.
\newblock \emph{\bibinfo{journal}{Science}} \textbf{\bibinfo{volume}{354}},
  \bibinfo{pages}{83--88} (\bibinfo{year}{2016}).

\bibitem{Gunther2008}
\bibinfo{author}{G{\"u}nther, U.} \& \bibinfo{author}{Samsonov, B.~F.}
\newblock \bibinfo{title}{Naimark-dilated p t-symmetric brachistochrone}.
\newblock \emph{\bibinfo{journal}{Phys. Rev. Lett.}}
  \textbf{\bibinfo{volume}{101}}, \bibinfo{pages}{230404}
  (\bibinfo{year}{2008}).

\bibitem{Kawabata2017}
\bibinfo{author}{Kawabata, K.}, \bibinfo{author}{Ashida, Y.} \&
  \bibinfo{author}{Ueda, M.}
\newblock \bibinfo{title}{Information retrieval and criticality in
  parity-time-symmetric systems}.
\newblock \emph{\bibinfo{journal}{Phys. Rev. Lett.}}
  \textbf{\bibinfo{volume}{119}}, \bibinfo{pages}{190401}
  (\bibinfo{year}{2017}).

\bibitem{Noggle1971}
\bibinfo{author}{Schirmer, R.~E.} \& \bibinfo{author}{Noggle, J.~H.}
\newblock \emph{\bibinfo{title}{The nuclear Overhauser effect; chemical
  applications by Joseph N. Noggle and Roger E. Schirmer}}
  (\bibinfo{publisher}{Academic Press}, \bibinfo{year}{1971}).

\end{thebibliography}
\end{document}